\documentclass[12pt]{article}
\usepackage{amsmath,graphicx,color,setspace}
\usepackage{subfigure,ametsoc}
\begin{document}

\title{Statistics of the seasonal cycle of the 1951-2000 surface temperature records in Italy}
\author{Valerio Lucarini\\Dipartimento di Matematica ed Informatica, Universit\'{a} di
Camerino
\\Via Madonna delle Carceri, 62032
Camerino (MC), Italy\\
\\ Teresa Nanni \\ISAC-CNR, Via Piero Gobetti 101, 40129 Bologna,
Italy \\ \\Antonio Speranza\\Dipartimento di Matematica ed
Informatica, Universit\'{a} di Camerino\\ Via Madonna delle
Carceri, 62032 Camerino (MC), Italy\\}

%
%
%

\maketitle

\newpage

\begin{abstract}
We present an analysis of seasonal cycle of the last 50 years of records of surface
temperature in Italy. We consider two data sets which synthesize
the surface temperature fields of Northern and Southern Italy.
Such data sets consist of records of daily maximum and minimum
temperature. We compute the best estimate of the seasonal cycle of
the variables considered by adopting the cyclograms' technique. We
observe that in general the minimum temperature cycle lags behind
the maximum temperature cycle, and that the cycles of the Southern
Italy temperatures records lag behind the corresponding cycles
referring to Northern Italy. All seasonal cycles lag considerably
behind the solar cycle. The amplitude and phase of the seasonal
cycles do not show any statistically significant trend in the time
interval considered.

\end{abstract}

\section{Introduction}
The analysis of the seasonal cycle of temperature records is of
the uttermost importance in order to provide a detailed
description of the climate of the geographical area under
consideration. A correct approach to the evaluation of the
seasonal signal allows to have a clearer picture of changes in
such a signal and at the same time permits a more precise position
of the problem of estimating the statistical properties, in terms
of short-time variability, long-term trend, and extremes, of the
residual signal.

In particular, the possibility of capturing with greater detail
the properties of the seasonal signal is especially relevant for
the analysis of regions, like the Mediterranean area, that are
characterized by relevant intermittence. The presence of
noticeable year-to-year variations for the seasonal cycle in the
Italian peninsula has been observed and reported in some of the
most relevant treatises of the past, from Roman Age - in Plinius
the Old's \textit{Naturalis Historia} - to early XIX century - in
Leopardi's \textit{Zibaldone}.

In this study we analyze the seasonal cycle for a 50-year period
(1951-2000) of the maximum and minimum temperature records of two
synthetic stations series, which synthesize the information
regarding Northern and Southern Italy. The data have been derived
from daily observations of temperatures taken in 64 stations
covering the Italian peninsula.

In order to provide the statistical description of the seasonal
signal of any record, able to quantify the \textit{mean} seasonal
cycle and as well as the properties of its \textit{short}- and
\textit{long-term variability}, we must have several, well defined
sampled estimates of its fundamental characteristics, namely
\textit{phase} and \textit{amplitude}.

In these terms, the application of the Discrete Fourier Transform
(henceforth, DFT) on the entire record is of relatively little
use, since it provides only the best - in terms of fraction of the
total variance - \textit{global} estimate of the phase and
amplitude of the $1 y^{-1}$ frequency component, while no
information is given on the variability of the seasonal cycle.

For each record, we estimate the seasonal component throughout the
record by considering the collection of all the \textit{local} (in
time) best estimate of seasonal cycle. Such an approach is along
the lines of the statistical technique proposed when introducing
the cyclograms \citep{Attolini1984,Galli1988}. The resulting
seasonal signal is not precisely periodic, since the phase and the
amplitude of the computed sinusoidal curve are not constant.
Therefore it is possible to statistically analyze how the
amplitude and phase of the seasonal signal vary with time.

Such an approach is viable because our data obey with the narrow
band approximation, \textit{i.e.} in each subset of the data used
for the local estimates, the spectrum of the data has a sharp,
narrow peak for the $1 y ^{-1}$ frequency component, so that the
phase and amplitude of the seasonal cycle are well defined.


We wish underline that very recently sophisticated DFT-based
techniques, which follow a different approach than ours, have been
proposed to assess simultaneously the diurnal, seasonal and
long-term variability of climate records \citep{Vinnikov2003}.

Our paper is organized as follows. In section \ref{data} we
describe the data sets considered in this work. In section
\ref{cyclo} we describe how it is possible to analyze a given
frequency component of a signal by considering a collection of its
local estimates. In section \ref{seasonal} we present the analysis
of the seasonal cycles of the data. In section \ref{trend} we
present an analysis of the significativity of the trends of the
estimated phase and amplitude of the seasonal signals. In section
\ref{deseason} we present the analysis of the de-seasonalized
data. In section \ref{concl} we present our conclusions.

\section{Data description}\label{data}
The data used in this study are derived from a set of station
records with daily minimum and maximum temperature observations
for a 50-year period (1951-2000). They were extracted from the
Italian Air Force (Aeronautica Militare, henceforth AM) climatic
database, that was recently used for the study of Italian daily
precipitation \citep{Brunetti2001,Brunetti2002}; cloud cover
\citep{Maugeri2001} and sea level pressure \citep{Maugeri2003} as
well. The AM climatic database includes 164 stations. Some of
them, however, cover only rather short periods, other ones have a
large number of missing data. Since we are interested in providing
information on the Italian climatology, we have selected a subset
of the stations which give a reasonable coverage of Italy and
which are provided with long and reliable records. The result was
a subset of 64 stations. The selected records were quality-checked
and in order to increase the confidence of the results,
homogenization was based, not only on AM records, but also on
records derived from other data sources such as Ufficio Centrale
di Ecologia Agraria, Servizio Idrografico, and some specific
research project that allowed daily series to be recovered for
several of the most important Italian observatories.

\begin{figure}[t]
\centering
 \subfigure[Station N maximum temperature]
   {\includegraphics[angle=270,width=0.45\textwidth]{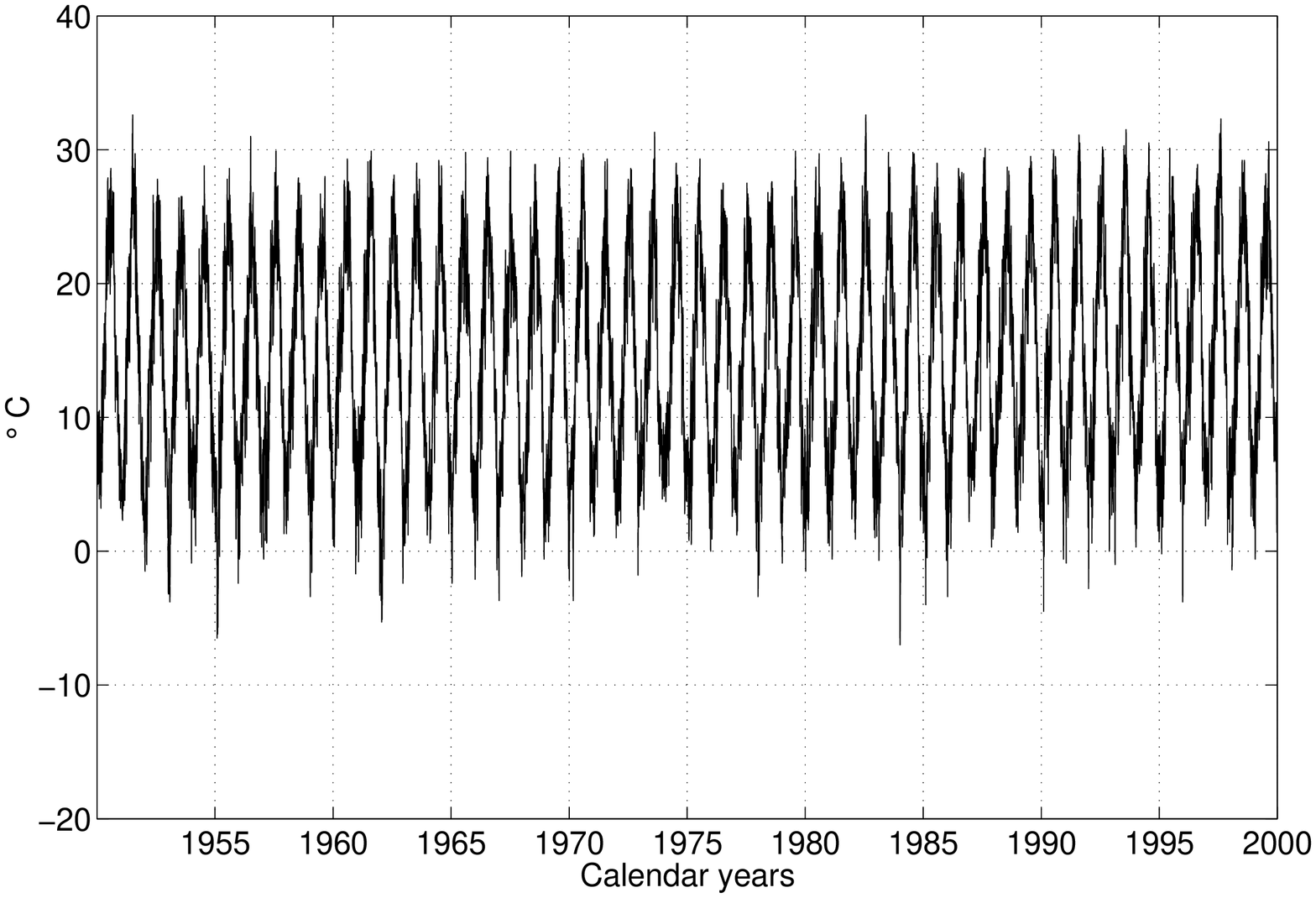}}
 \hspace{5mm}
 \subfigure[Station S maximum temperature]
   {\includegraphics[angle=270,width=0.45\textwidth]{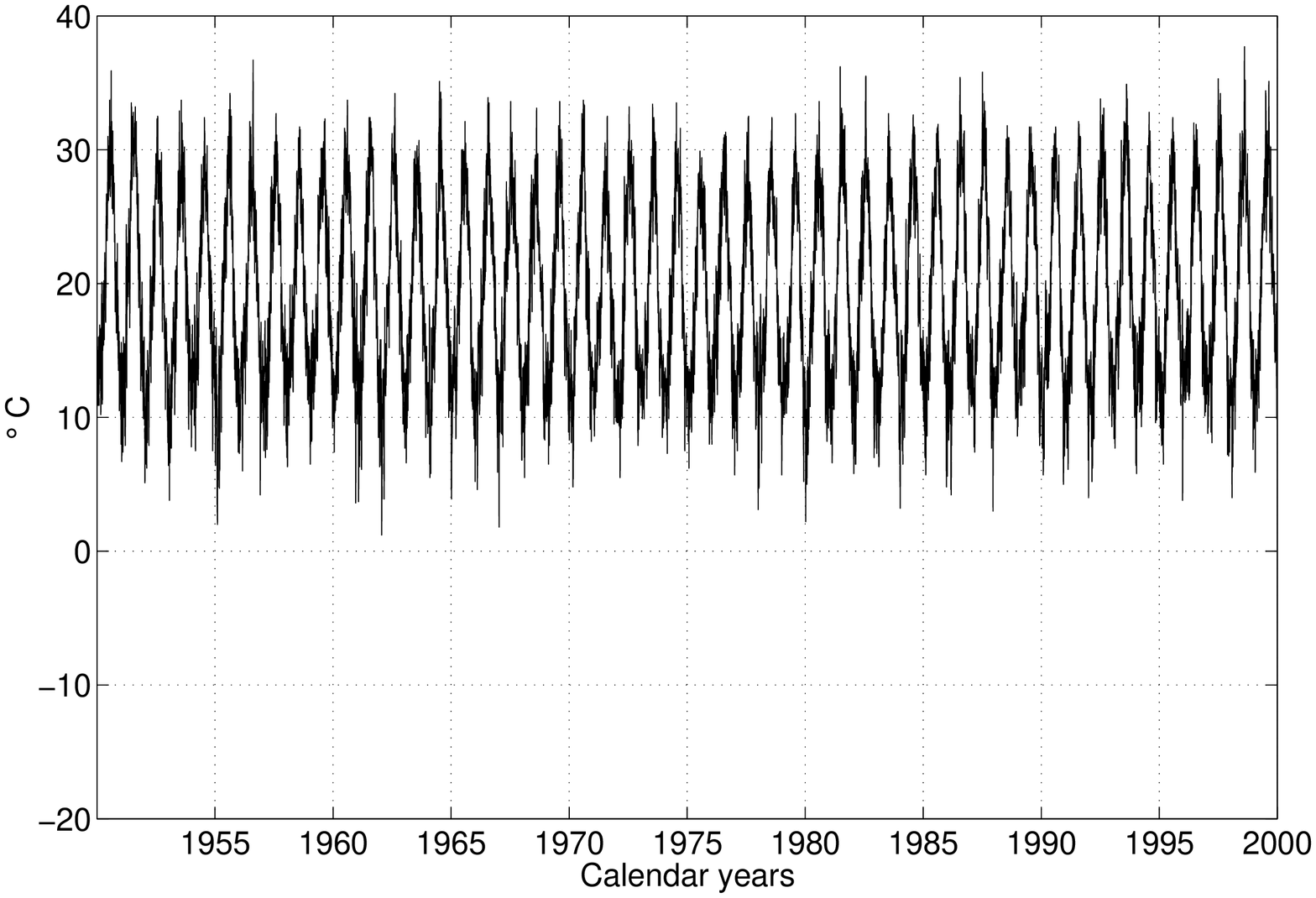}}\\
 \subfigure[Station N minimum temperature]
   {\includegraphics[angle=270,width=0.45\textwidth]{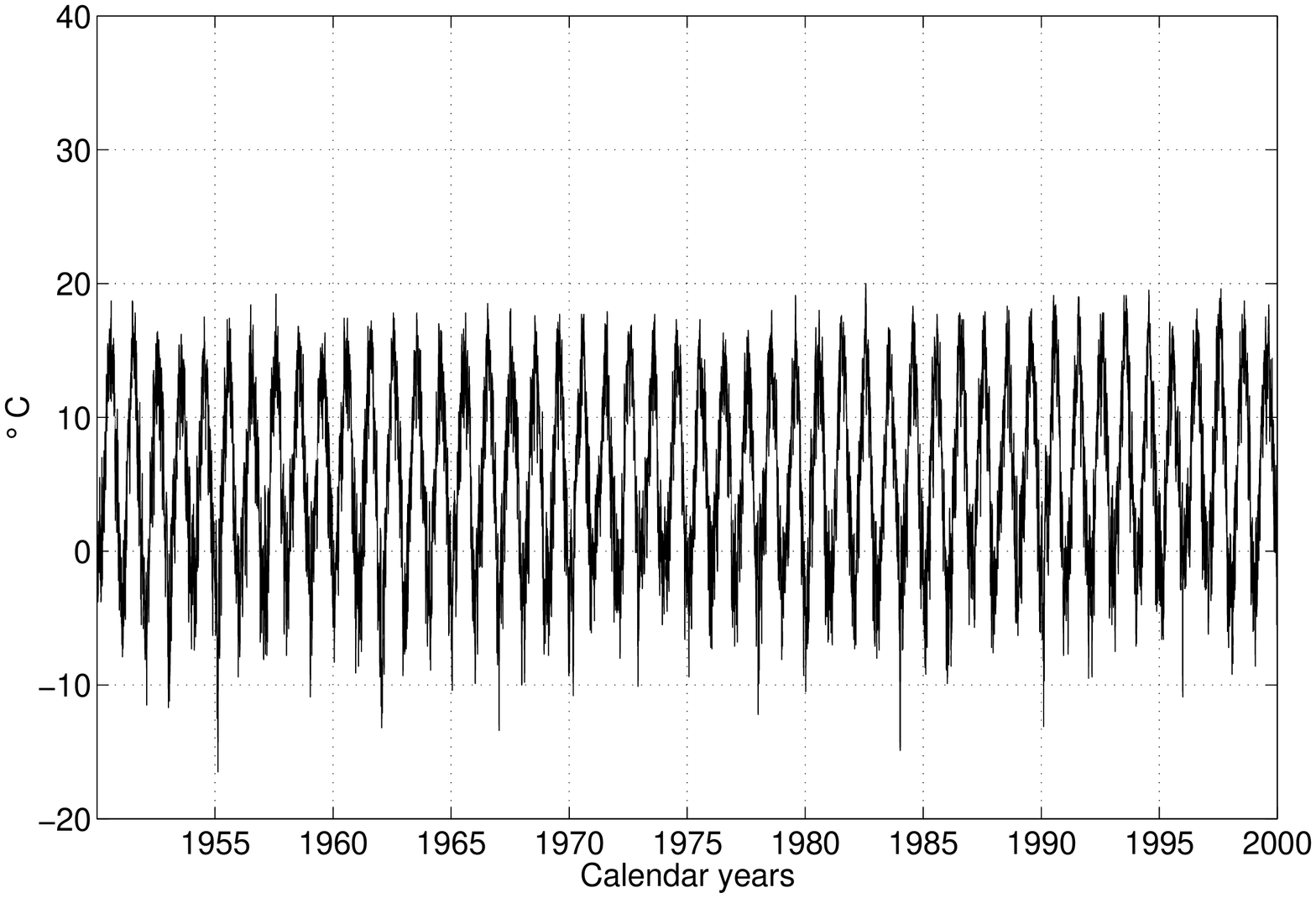}}
 \hspace{5mm}
 \subfigure[Station S minimum temperature]
   {\includegraphics[angle=270,width=0.45\textwidth]{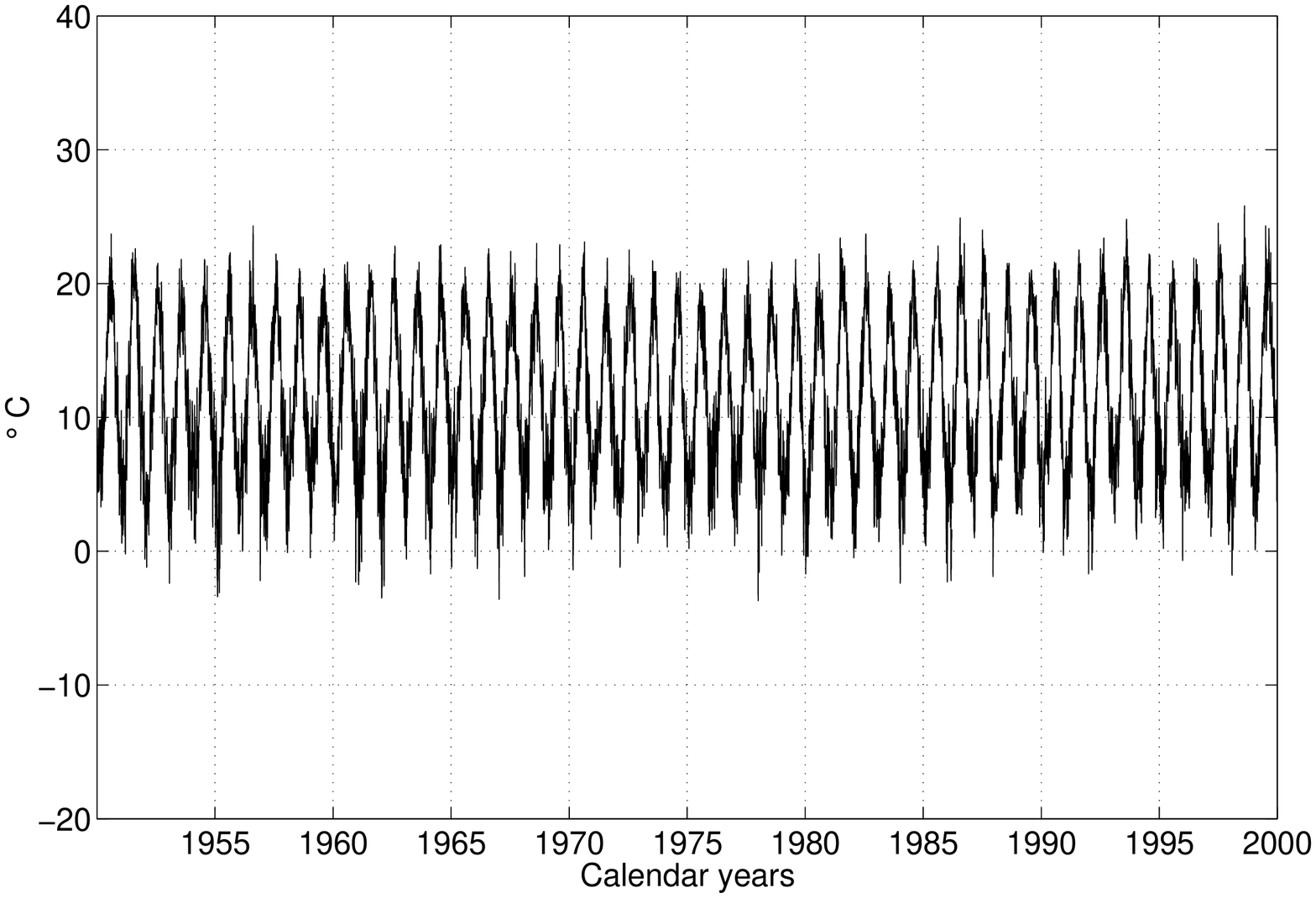}}
 \caption{Maximum and minimum temperature records of station N and station S.}\label{datafigure}
 \end{figure}

EOF analysis, which will be fully reported in a future
publication, shows that the daily maximum and minimum temperature
data fields can be reduced with a good degree of approximation to
two degrees of freedom. In both cases, these degrees of freedom
contribute to over $90\%$ of the variance of the signal.
The first two principal components are representative of the two
geographically distinct areas of Northern and Southern Italy.
Therefore, it has been possible to create two synthetic data sets
for Northern and Southern Italy, which henceforth we refer to as
station N and station S temperature records, respectively. Each of
the 64 stations has been assigned to either station N or station S
on the basis of a score. Then the station N and station S
synthetic data sets have been created by suitably averaging the
data of the corresponding stations. Each resulting data set
consists of the records of daily maximum and minimum temperature,
which are henceforth indicated as $T_{max}^{N/S}$ and
$T_{min}^{N/S}$, with obvious meaning of the indexes. These data are depicted in figure \ref{datafigure}.
Qualitatively, the geographic boundary dividing the stations
contributing respectively to the station N and S data sets is
along the parallel between Firenze (Tuscany) and Bologna (Emilia
Romagna).

\section{Local estimate of a given frequency component}\label{cyclo}

We consider the statistical approach related to the technique of
cyclograms \citep{Attolini1984,Galli1988}. Such an approach
provides the possibility of capturing the amplitude and phase
time-dependent variations of a given frequency sine wave component
of the signal under examination \cite{Attolini1985,Attolini1989}.

Given a signal $x\left(t\right)$, $t=1,\ldots,N$, a frequency
$2\pi/\tau$ and a time window $2T+1$, we consider the centered
moving average over $2T+1$ terms of the series
$\left\{x\left(t\right)\exp\left[-\textrm{i}2\pi t
/\tau\right]\right\}$:
\begin{equation}\label{cyclo1}
a\left(t;\tau,T\right)=\frac{1}{2T+1}\sum_{j=t-T}^{t+T}x\left(j\right)\exp\left[-\textrm{i}2\pi
j /\tau\right],
\end{equation}
where $T+1\leq t\leq N-T$ since the signal has $N$ samplings.

If the frequency $2\pi/\tau$ is an integer multiple of $2\pi/N$,
we have that $a\left(t;\tau,T\right)$ can be expressed as the DFT
of a suitably convolution product:
\begin{equation}\label{cyclo2}
a\left(t;\tau,T\right)=\frac{N}{2T+1}DFT\left[x\ast
w\right]\left(2\pi/\tau\right)
\end{equation}
where the first factor is a renormalization constant, $\ast$
represents the convolution product, and $w$ is the weighting
function:
\begin{equation}\label{weight}
w\left(t\right)=
\begin{cases}
\frac{1}{2T+1}, \hspace{4mm} 1\leq t \leq T+1 \\
0, \hspace{4mm} T+2\leq t \leq N-T\\
\frac{1}{2T+1}, \hspace{4mm} N-T+1\leq t \leq N \\
\end{cases}
\end{equation}
Equations (\ref{cyclo1}-\ref{cyclo2}) imply that, if $2\pi/\tau$
belongs to the discrete spectrum of the signal, and if $2T+1\geq
\tau$, $a\left(t;\tau,T\right)$ is related to the best estimate of
the $2\pi/\tau$ frequency sine $S\left(t,2\pi/\tau\right)$ and
cosine $C\left(t,2\pi/\tau\right)$ wave components of the portion
$t-T\leq t \leq t+T$ of the signal $x\left(t\right)$ as follows:
\begin{align}\label{cyclo3}
C\left(t,2\pi/\tau\right)=&\frac{2}{2T+1}\textrm{Re}\left[a\left(t;\tau,T\right)\right]\\
S\left(t,2\pi/\tau\right)=&-\frac{2}{2T+1}\textrm{Im}\left[a\left(t;\tau,T\right)\right]
\end{align}
where $\textrm{Re}$ and $\textrm{Im}$ indicate the real and
imaginary part, respectively. Therefore, we can construct a
\textit{global} best estimate of the $2\pi/\tau$ frequency signal
$\Sigma\left(t,2\pi/\tau\right)$ for each value of $T+1\leq t \leq
N-T$ by considering all the \textit{local} best estimates obtained
using the result contained in equation (\ref{cyclo3}):
\begin{equation}\label{seasonal1}
\begin{split}
\Sigma\left(t,2\pi/\tau\right)&=C\left(t,2\pi/\tau\right)\cos\left(2\pi
t /\tau\right)+S\left(t,2\pi/\tau\right)\sin\left(2\pi t
/\tau\right)\\
&=A\left(t,2\pi/\tau\right)\cos\left(2\pi t
/\tau+\phi\left(t,2\pi/\tau \right) \right),
\end{split}
\end{equation}
where:
\begin{align}\label{amplitude}
A\left(t,2\pi/\tau\right)&=\sqrt{C\left(t,2\pi/\tau\right)^2+S\left(t,2\pi/\tau\right)^2},\\
\phi\left(t,2\pi/\tau
\right)&=-\arctan\left[\frac{S\left(t,2\pi/\tau\right)}{C\left(t,2\pi/\tau\right)}\right].
\end{align}
We can reasonably extend the function
$\Sigma\left(t,2\pi/\tau\right)$ to the whole range $t=1,\ldots,N$
in the following way:
\begin{equation}\label{seasonal2}
\overline{\Sigma}\left(t,2\pi/\tau\right)=\begin{cases}A\left(T+1,2\pi/\tau\right)\cos\left(2\pi
t /\tau+\phi\left(T+1,2\pi/\tau \right) \right), \hspace{4mm} t<T+1\\
\Sigma\left(t,2\pi/\tau\right), \hspace{4mm} T+1\leq t\leq N-T\\
A\left(N-T,2\pi/\tau\right)\cos\left(2\pi t
/\tau+\phi\left(N-T,2\pi/\tau \right) \right), \hspace{4mm} t>N-T
\end{cases}
\end{equation}
Since the coefficients of the sine and cosine waves change with
$t$, the signal $\overline{\Sigma}\left(t,2\pi/\tau\right)$ is not
purely periodic, \textit{i.e.} its DFT does not have $2\pi/\tau$
as only nonzero component. Obviously, the more persistent with $t$
are the phase and amplitude of the local estimates of the
$2\pi/\tau$ signal, the more monochromatic is
$\overline{\Sigma}\left(t,2\pi/\tau\right)$.

Phase cyclograms \citep{Attolini1984,Galli1988} provide a very
synthetic way of picturing the phase variations of the selected
frequency components of the signals. The $x-$ and $y-$ components
of the phase cyclogram of a signal can be constructed in the
following way:
\begin{align}
PHX\left(t,2\pi/\tau\right)&=\sum_{j=1}^{t}C\left(j,2\pi/\tau\right)/A\left(j,2\pi/\tau\right),\\
PHY\left(t,2\pi/\tau\right)&=\sum_{j=1}^{t}S\left(j,2\pi/\tau\right)/A\left(j,2\pi/\tau\right).
\end{align}
The more coherent in phase is the frequency component of the
signal under examination, the more similar is the resulting graph
to a straight line. In the limiting case of a purely periodic
signal we actually obtain a straight line, whose angle with the
horizontal axis is the phase of the signal, apart from a constant.

\section{Seasonal cycles}\label{seasonal}

In order to apply the techniques presented in the previous chapter
to the temperature records $T_{max}^{N}$, $T_{max}^{S}$,
$T_{min}^{N}$, and $T_{min}^{S}$ we have performed a light
preprocessing procedure to the data. First of all, the four
records presented few missing data, ranging from a minimum of 3
($T_{max}^{N}$) to a maximum of 5 ($T_{max}^{S}$). We have filled
the holes with simple linear interpolations. Moreover, in order to
homogenize the length of the years, we have suppressed the
additional data of February occurring in each of the 12 bissextile
years of the time frame considered. Since these corrections regard
in each case less than $0.1\%$ of the total record, we are
confident that this procedure does not alter relevantly the
results later presented.

Since we are interested in evaluating the seasonal cycle, we
consider in equation (\ref{seasonal1}) $\tau=\tau_{0}=365$. The
most natural time window suitable for having a local estimate of
the seasonal cycle is clearly one year as well. Therefore, we
select $2T+1=2T_{0}+1=\tau_0=365$. This choice for $2T+1$ implies
that, following equation (\ref{cyclo1}), we have $49 \times 365+1$
local estimates of the seasonal cycle.

It is important to underline that such an approach is sensitive
only if the signal obeys the narrow band approximation,
\textit{i.e.} the spectrum of the signal has a strong, narrow peak
for the annual cyclic component. If, on the contrary, the signal
were characterized by a broad spectral feature comprising the $1
y^{-1}$ frequency component, it would be a mathematical nonsense
to investigate whether the seasonal cycle is changing. In such a
case the seasonal cycle is just \textit{not defined}, because
several contiguous spectral components having different
frequencies and shifting phase differences give contributions of
comparable importance.

\begin{figure}[t]
\centering
 \subfigure[Station N maximum temperature]
   {\includegraphics[angle=270,width=0.45\textwidth]{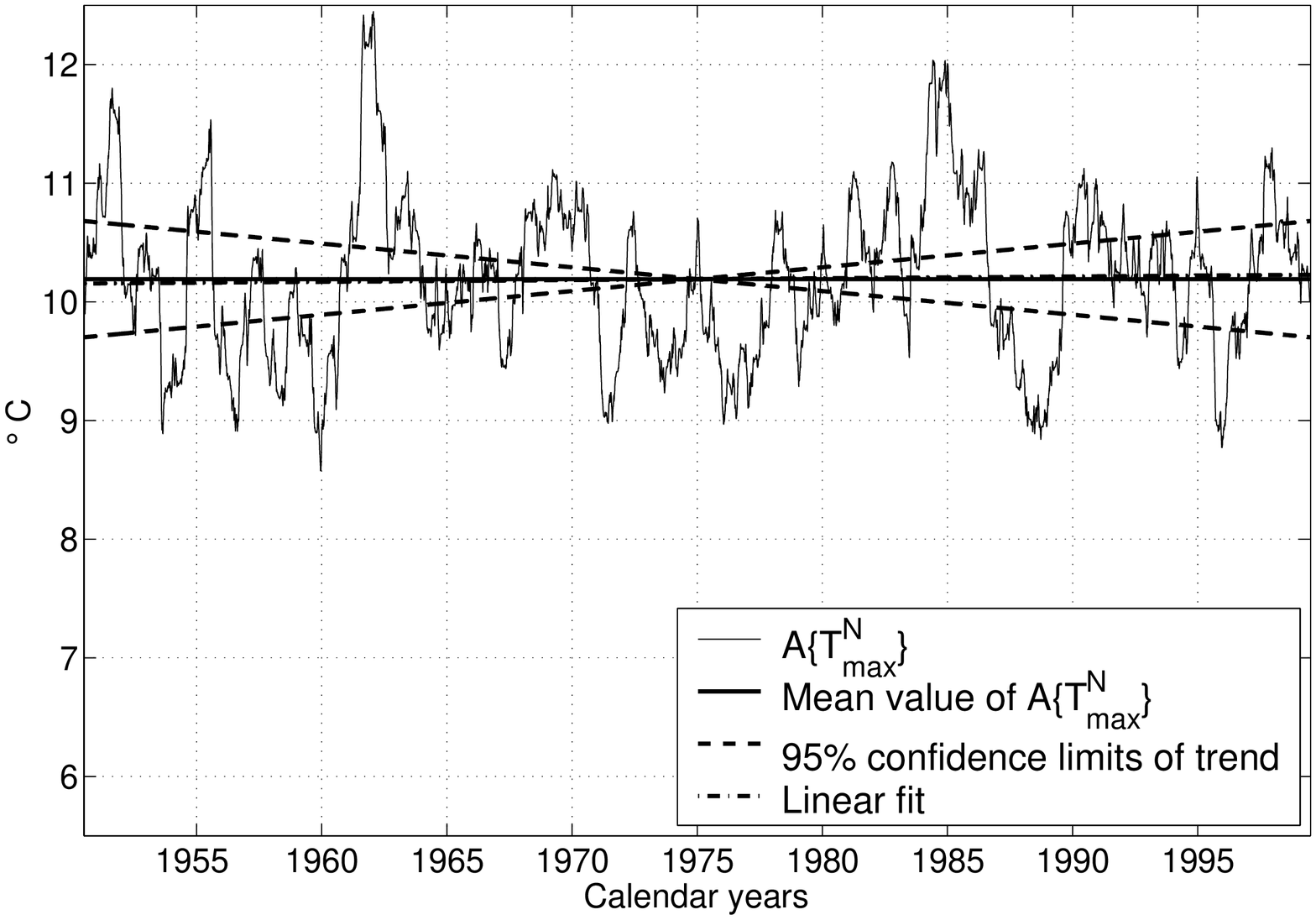}}
 \hspace{5mm}
 \subfigure[Station S maximum temperature]
   {\includegraphics[angle=270,width=0.45\textwidth]{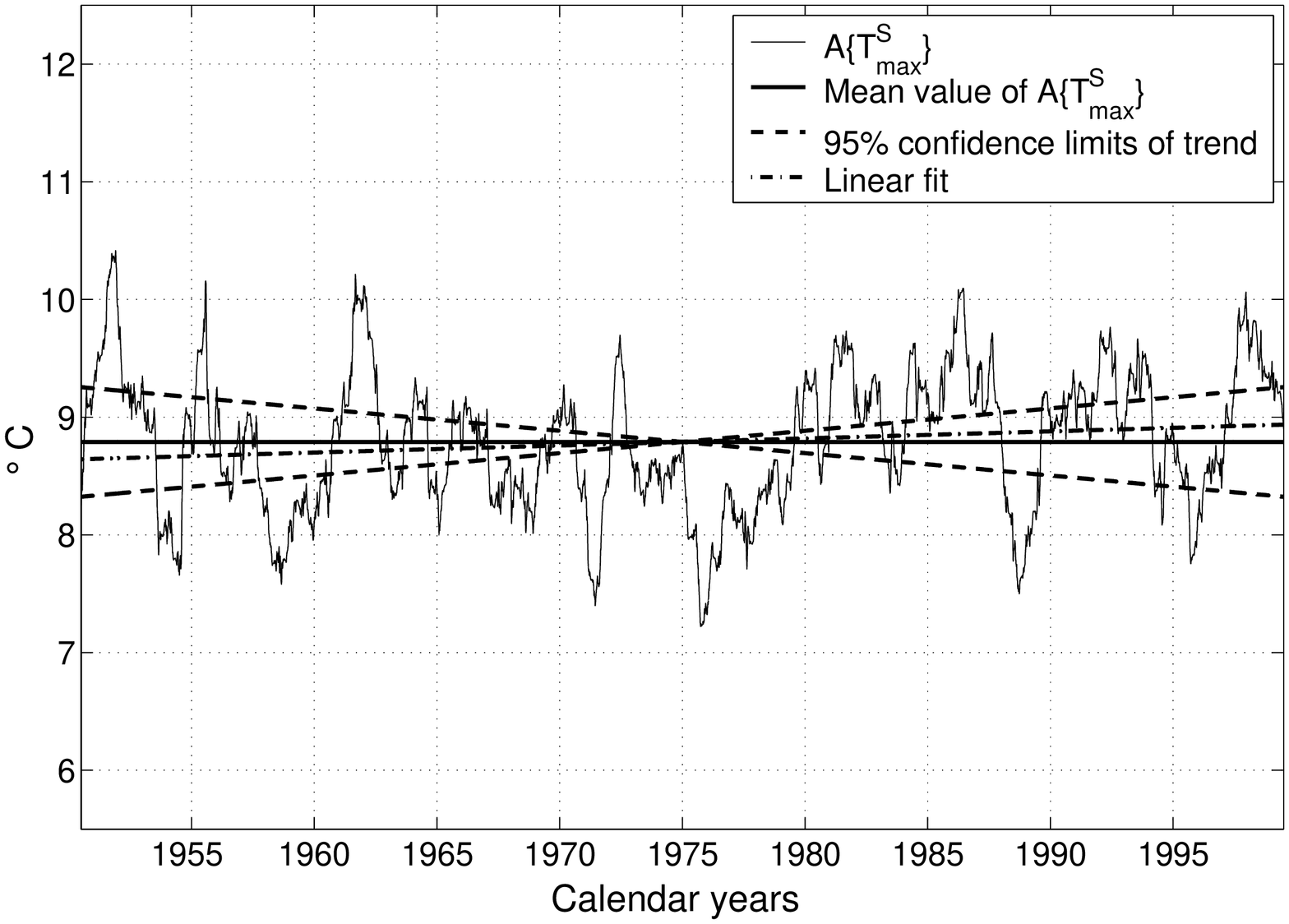}}\\
 \subfigure[Station N minimum temperature]
   {\includegraphics[angle=270,width=0.45\textwidth]{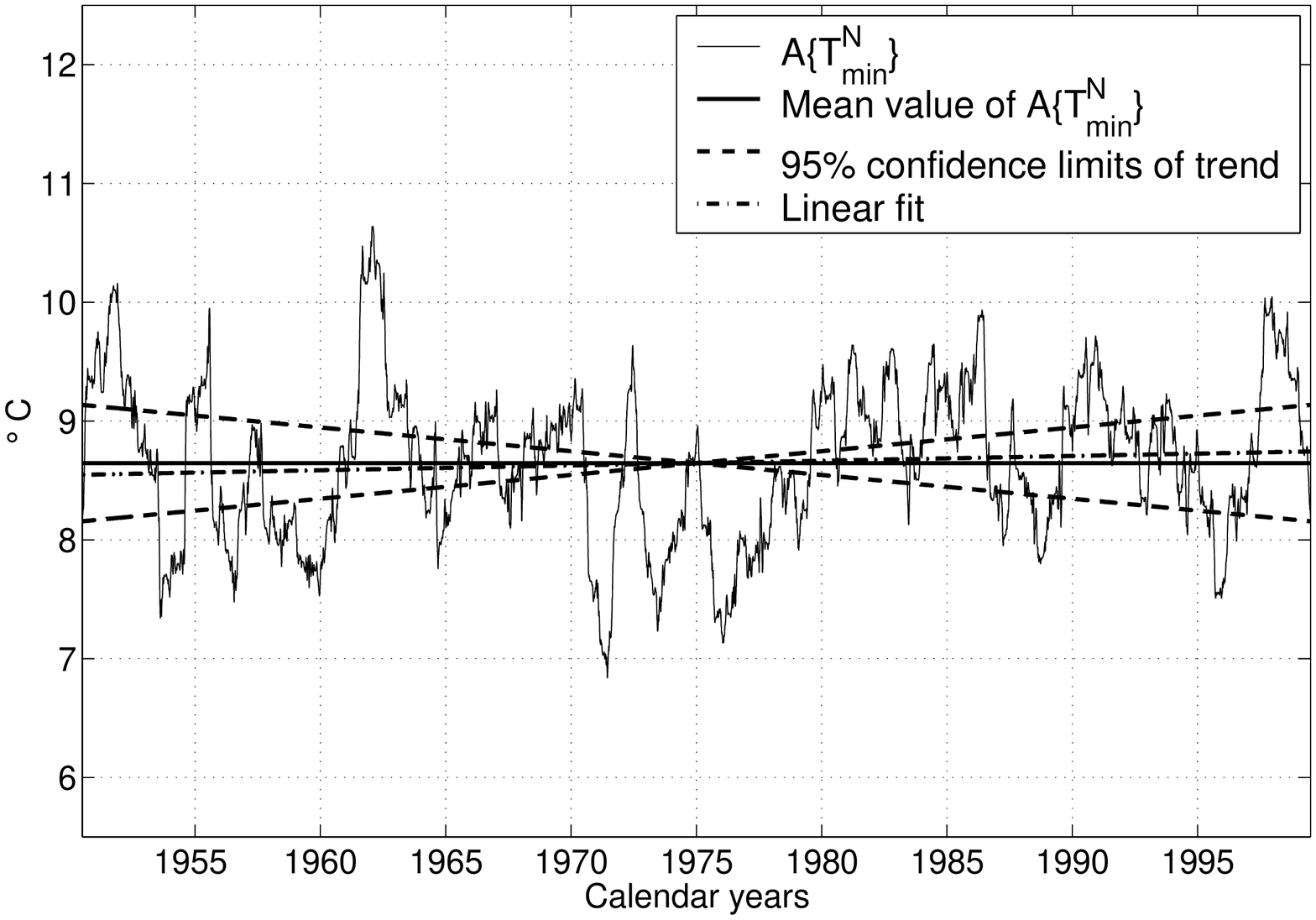}}
 \hspace{5mm}
 \subfigure[Station S minimum temperature]
   {\includegraphics[angle=270,width=0.45\textwidth]{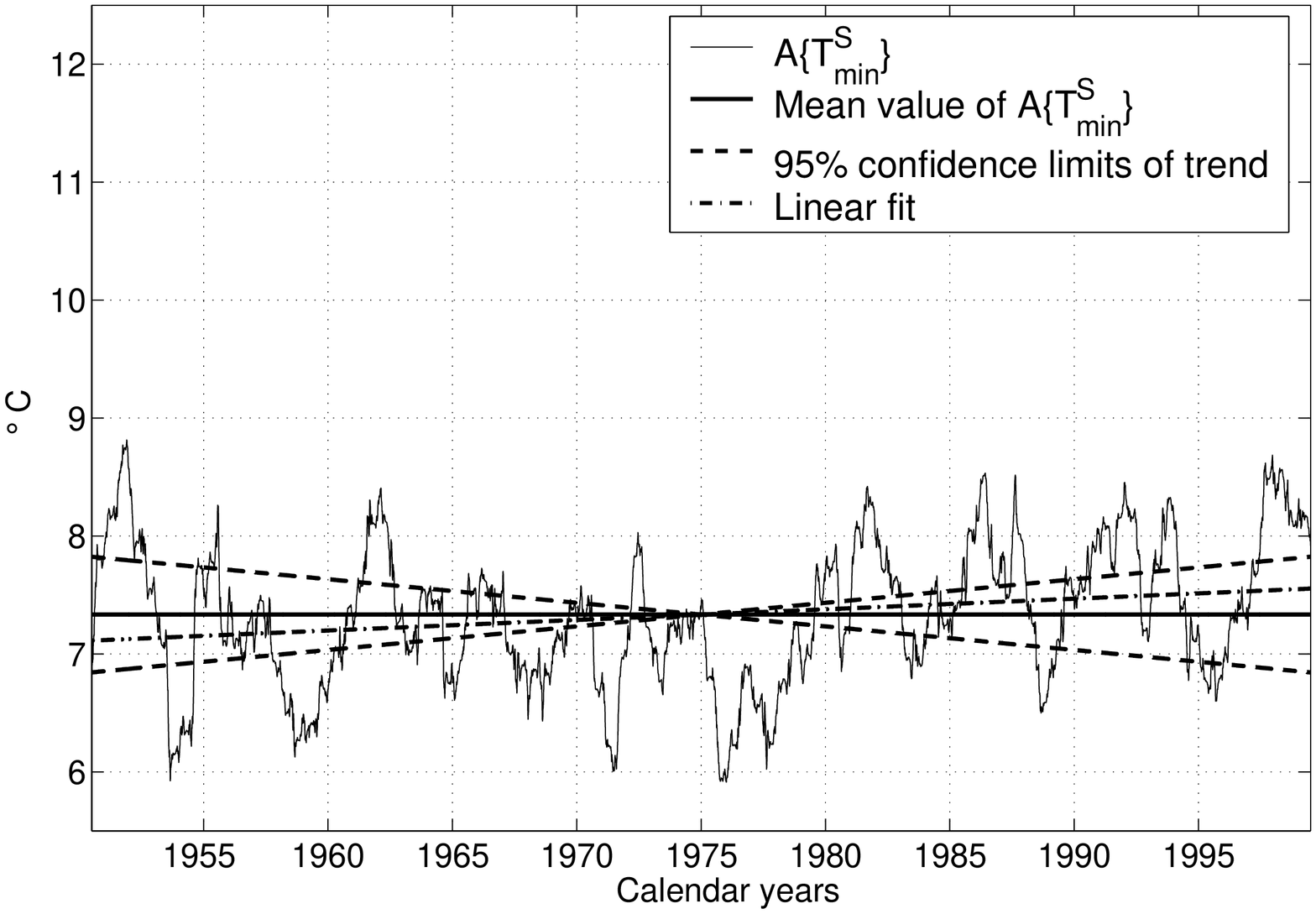}}
 \caption{Amplitude of the seasonal cycle of the maximum and minimum temperature records of station N and station S.}\label{ampfigure}
 \end{figure}

\begin{figure}[t]
\centering
 \subfigure[Station N maximum temperature]
   {\includegraphics[angle=270,width=0.45\textwidth]{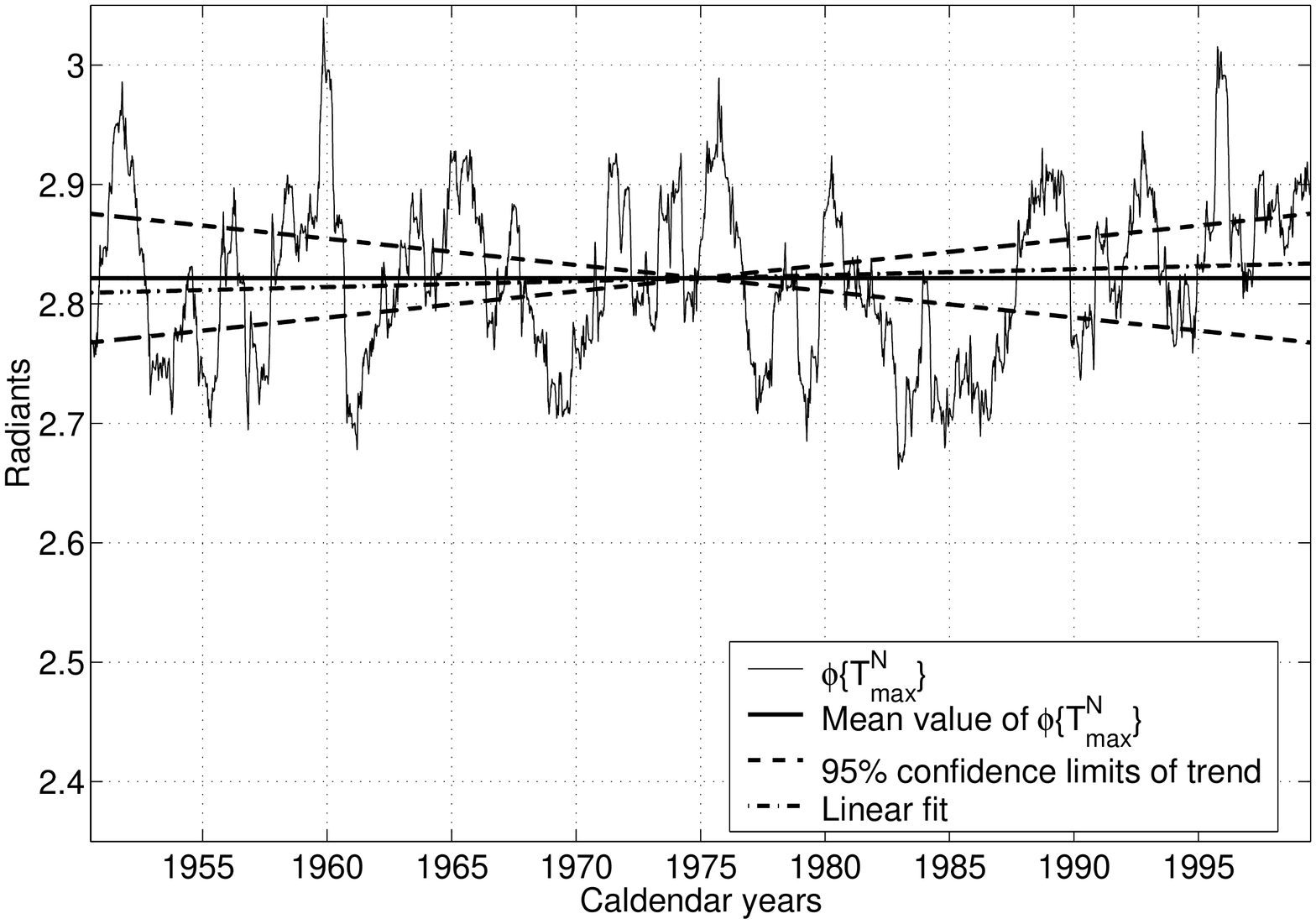}}
 \hspace{5mm}
 \subfigure[Station S maximum temperature]
   {\includegraphics[angle=270,width=0.45\textwidth]{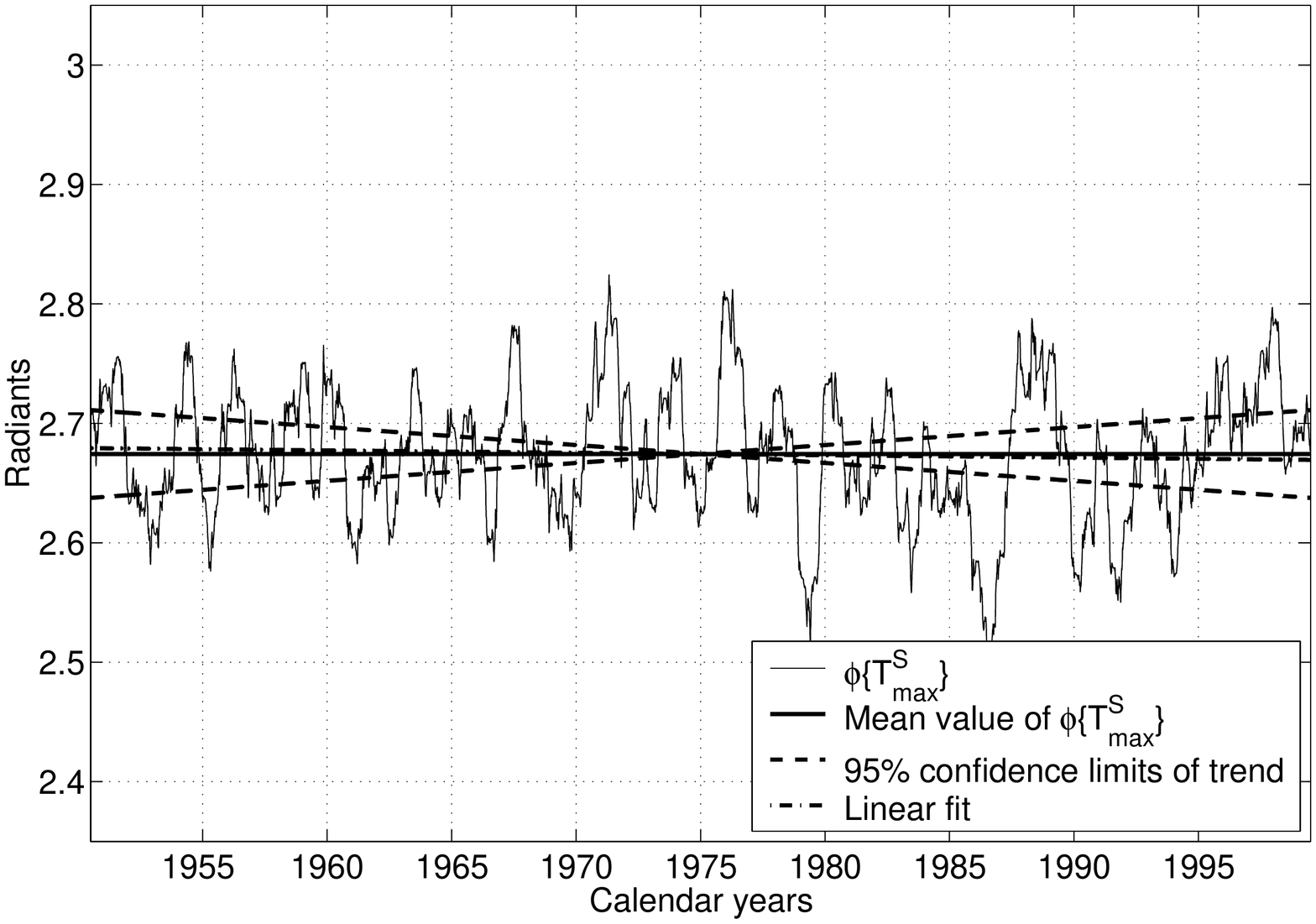}}\\
 \subfigure[Station N minimum temperature]
   {\includegraphics[angle=270,width=0.45\textwidth]{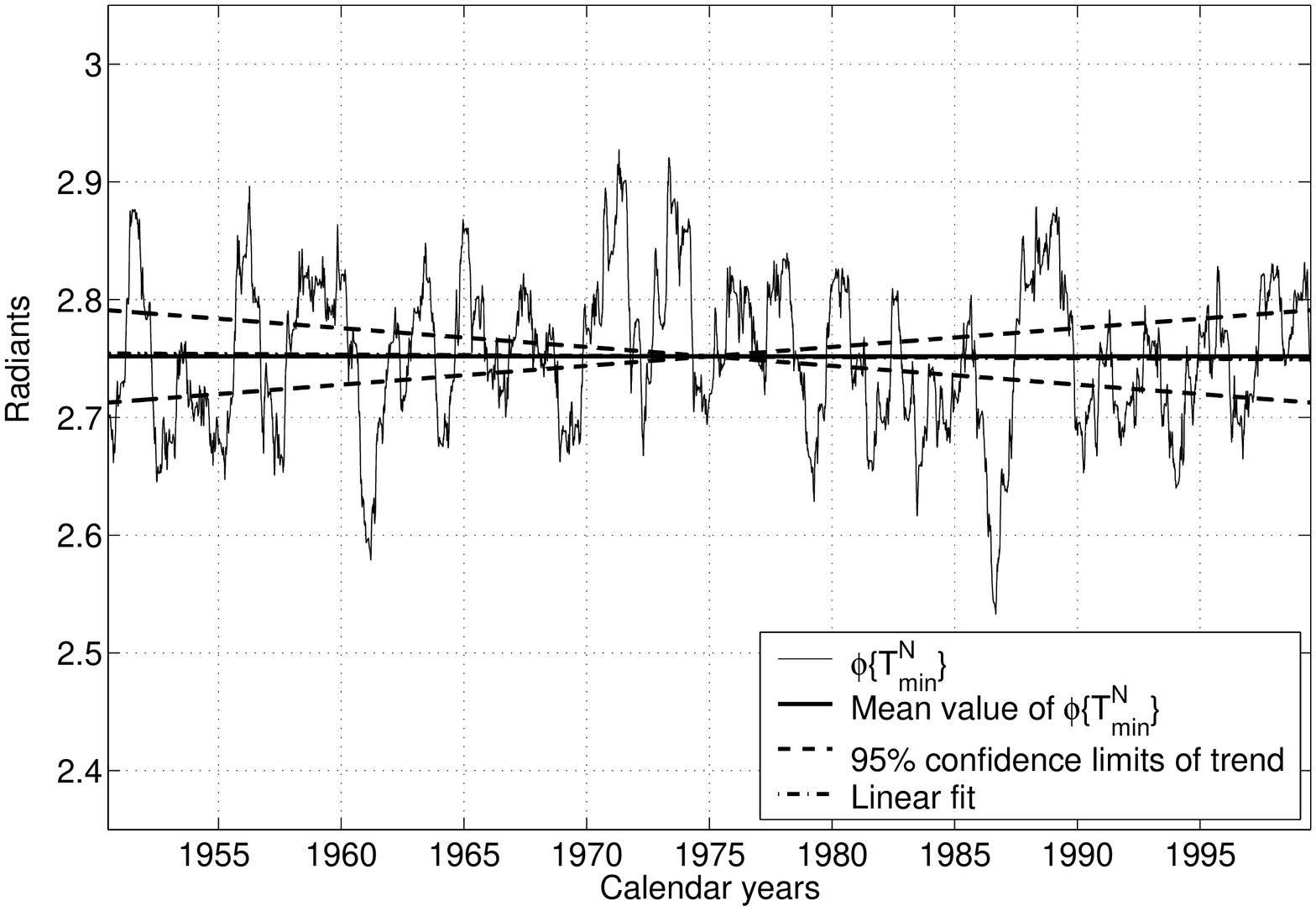}}
 \hspace{5mm}
 \subfigure[Station S minimum temperature]
   {\includegraphics[angle=270,width=0.45\textwidth]{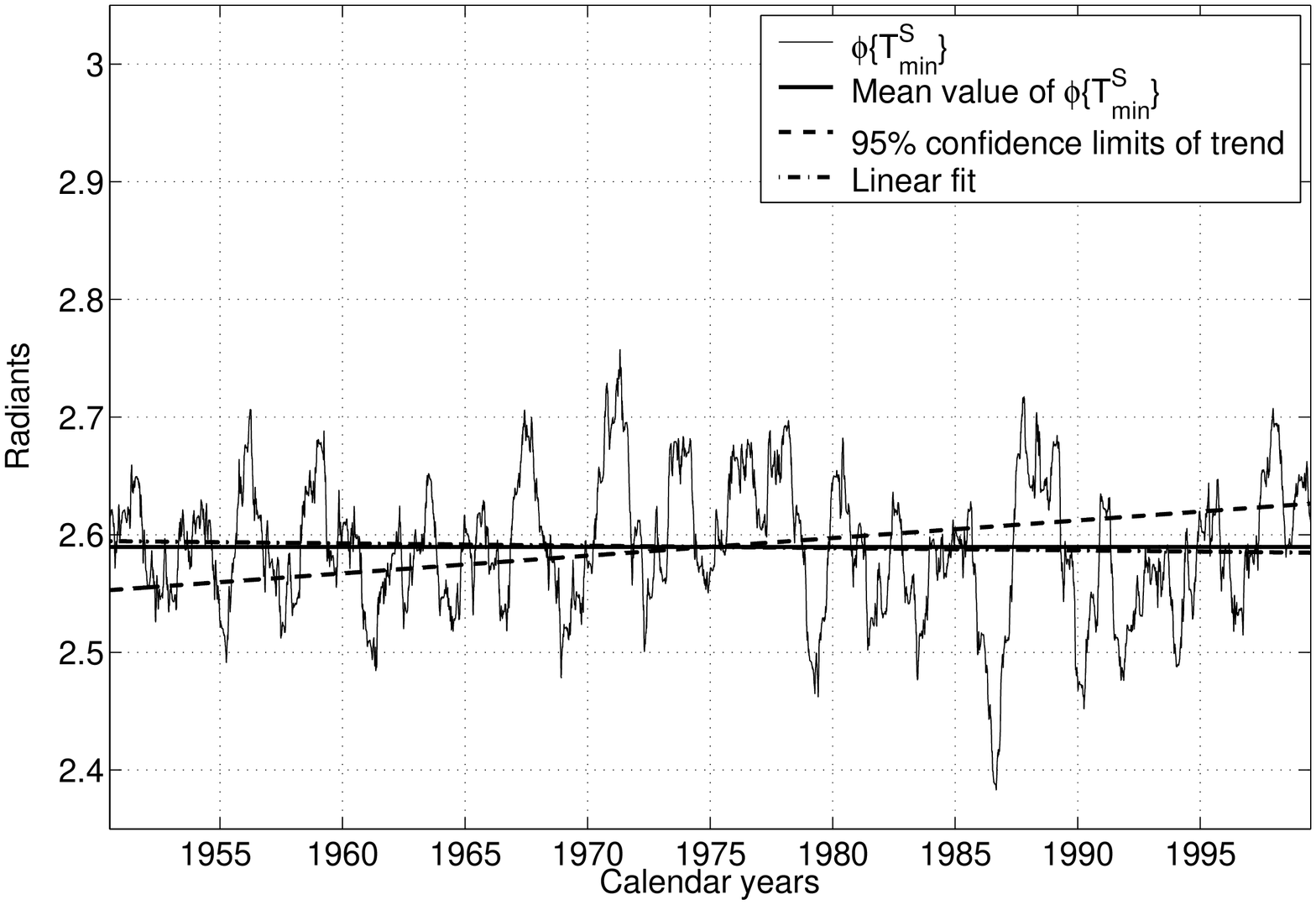}}
 \caption{Phase of the seasonal cycle of the maximum and minimum temperature records of station N and station S.}\label{phasefigure}
 \end{figure}

\begin{figure}[t]
\centering
 \subfigure[$\overline{\Sigma}\left(t,2\pi/\tau_0\right)\left\{T_{max}^{N}\right\}$]
   {\includegraphics[angle=270,width=0.45\textwidth]{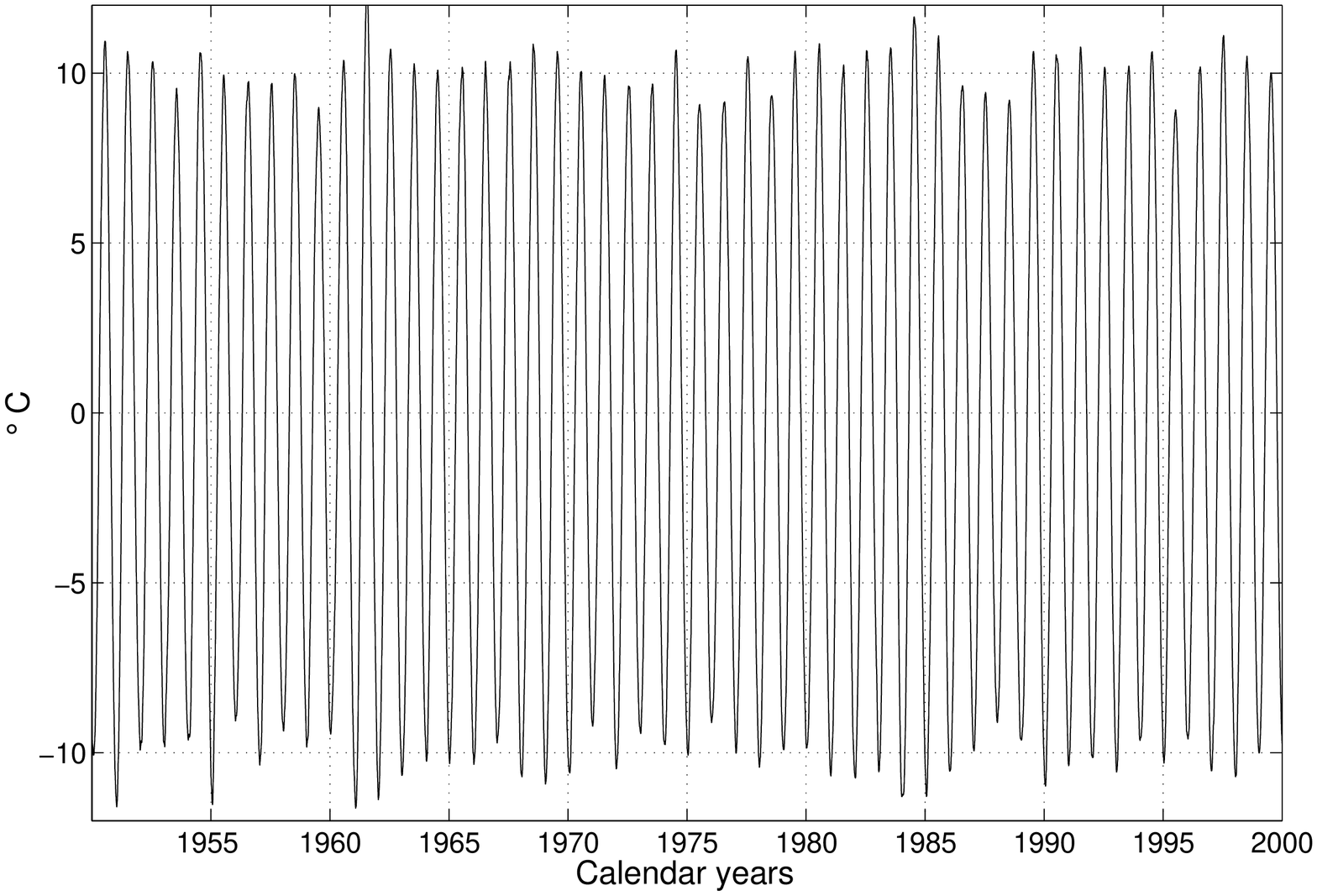}}
 \hspace{5mm}
 \subfigure[$\overline{\Sigma}\left(t,2\pi/\tau_0\right)\left\{T_{max}^{S}\right\}$]
   {\includegraphics[angle=270,width=0.45\textwidth]{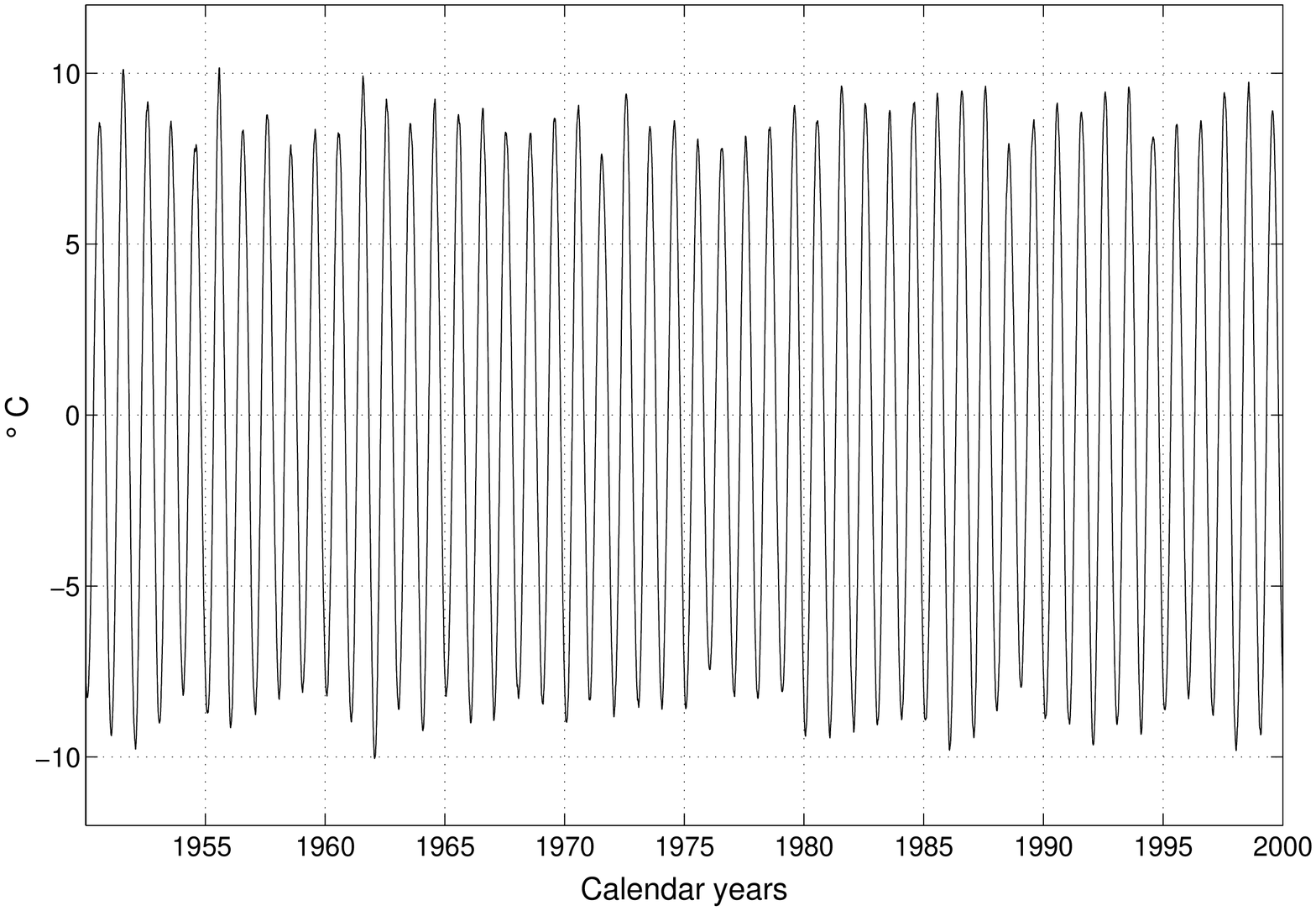}}\\
 \subfigure[$\overline{\Sigma}\left(t,2\pi/\tau_0\right)\left\{T_{min}^{N}\right\}$]
   {\includegraphics[angle=270,width=0.45\textwidth]{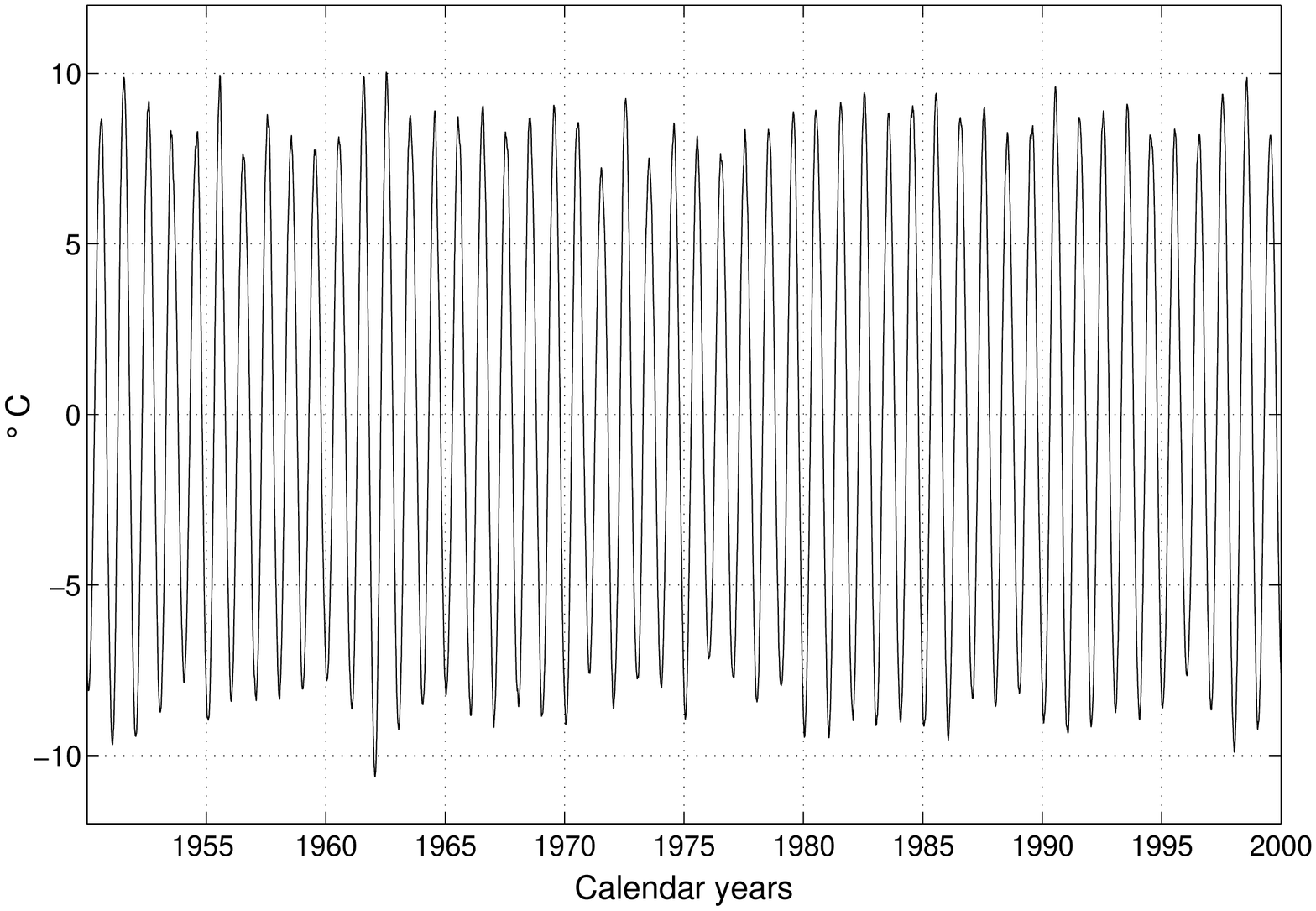}}
 \hspace{5mm}
 \subfigure[$\overline{\Sigma}\left(t,2\pi/\tau_0\right)\left\{T_{min}^{S}\right\}$]
   {\includegraphics[angle=270,width=0.45\textwidth]{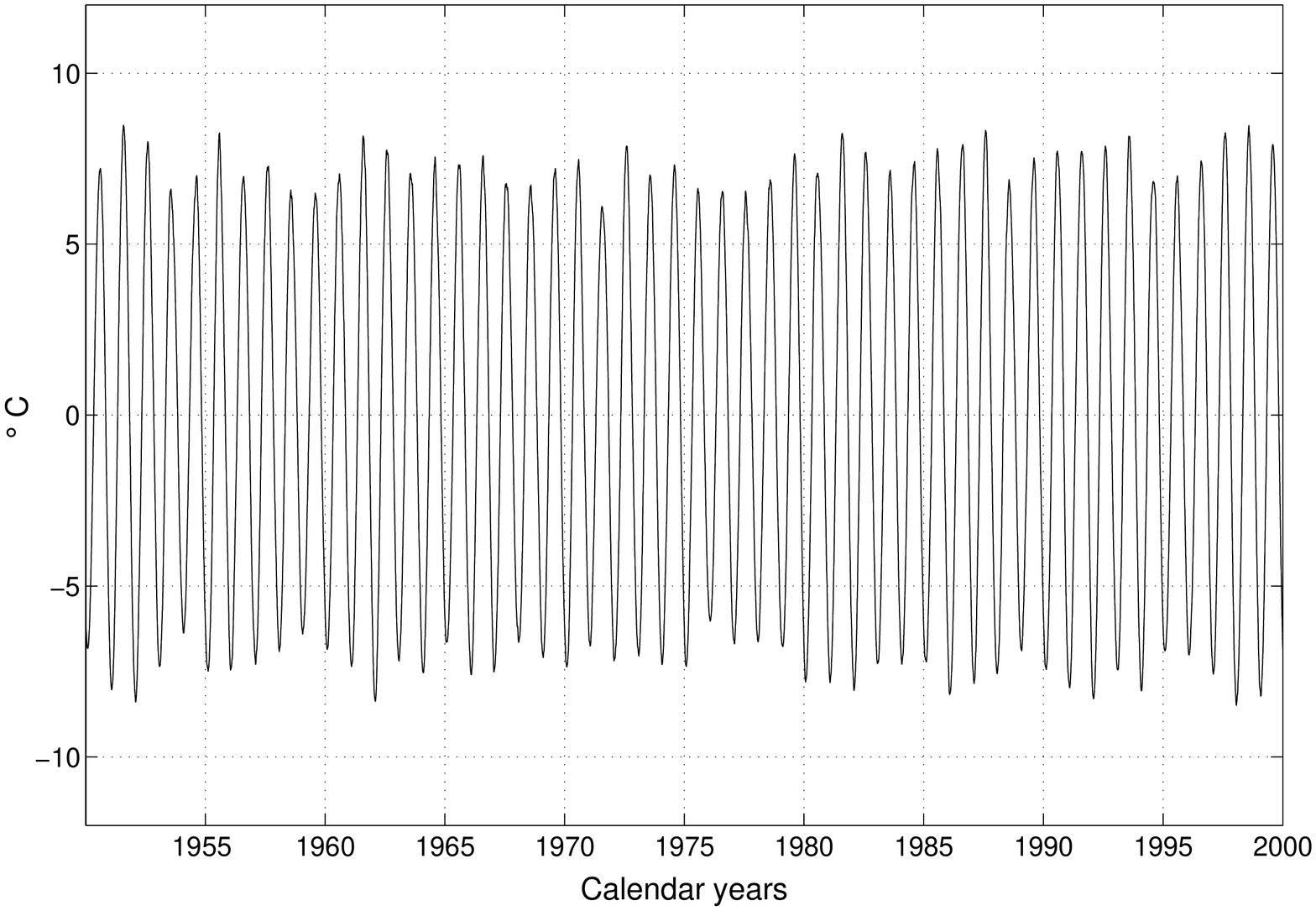}}
 \caption{Seasonal cycle of the maximum and minimum temperature records of station N and station S.}\label{cyclefigure}
 \end{figure}

The results we obtain for the amplitude signals are summarized in
table \ref{amptable} and depicted in figure \ref{ampfigure}, while
the results referring to the phase signals are reported in table
\ref{phasetable} and depicted in figure \ref{phasefigure}. In
figure \ref{cyclefigure} we present the results obtained for the
function $\overline{\Sigma}\left(t,2\pi/\tau_0\right)$ for the
four records considered.

The first result we want to point out is that there is no
statistically significant linear trend in either the amplitude of
the phase of the seasonal signal. In other terms, our analysis
suggests that in Italy in the time frame 1951-2000 \textit{seasons
have not changed} in their annual evolution. The statistical
analysis of the trend of the signals is described in detail in a
later subsection. We underline that in general it is sensible to
perform the analysis of the time-dependence of the seasonal signal
properties only if the record comprises several seasonal cycles.
In our case such condition is obeyed, since we have $N \gg
2T_0+1$.

The second result we wish to emphasize is that the amplitude of
the seasonal signal is significantly larger for maximum than for
minimum temperature, and that is significantly larger for
variables referring to Northern Italy. Moreover, the two effects
roughly sum up linearly, \textit{i.e.}:
\begin{equation}
\left\langle A\left\{T_{max}^{N}\right\}\right\rangle-\left\langle
A\left\{T_{max}^{S}\right\}\right\rangle \approx \left \langle
A\left\{T_{min}^{N}\right\}\right\rangle-\left\langle
A\left\{T_{min}^{S}\right\}\right\rangle,
\end{equation}
where we have dropped the $t$- and $\tau$-dependencies of $A$ for
sake of simplicity and where the notation $\langle \rangle$
indicates the mean value. Another interesting result is that for
both N and S stations the seasonal signal of minimum temperature
has an average phase delay with respect to the seasonal signal of
the maximum temperature. Moreover, the seasonal cycle of the
temperature records of station S has a delay with respect to the
seasonal cycle of the corresponding temperature records of station
N. Also in this case the two effect roughly sum up linearly:
\begin{align}
\left\langle
\phi\left\{T_{max}^{N}\right\}\right\rangle-\left\langle
\phi\left\{T_{max}^{S}\right\}\right\rangle &\approx \left \langle
\phi\left\{T_{min}^{N}\right\}\right\rangle-\left\langle
\phi\left\{T_{min}^{S}\right\}\right\rangle \approx 0.15 \approx 9
d \\
\left\langle
\phi\left\{T_{max}^{N}\right\}\right\rangle-\left\langle
\phi\left\{T_{min}^{N}\right\}\right\rangle &\approx \left \langle
\phi\left\{T_{max}^{S}\right\}\right\rangle-\left\langle
\phi\left\{T_{min}^{S}\right\}\right\rangle \approx 0.7 \approx
4d,
\end{align}
where we have expressed the phase differences in terms of calendar
days $d$. The maximum temperature record of station N is the
closest in terms of phase delay to the solar cycle, which
constitutes a fundamental forcing to the system. Such delay
corresponds to $\approx 30 d$.

\begin{figure}[t]
\centering
\includegraphics[angle=270,width=0.9\textwidth]{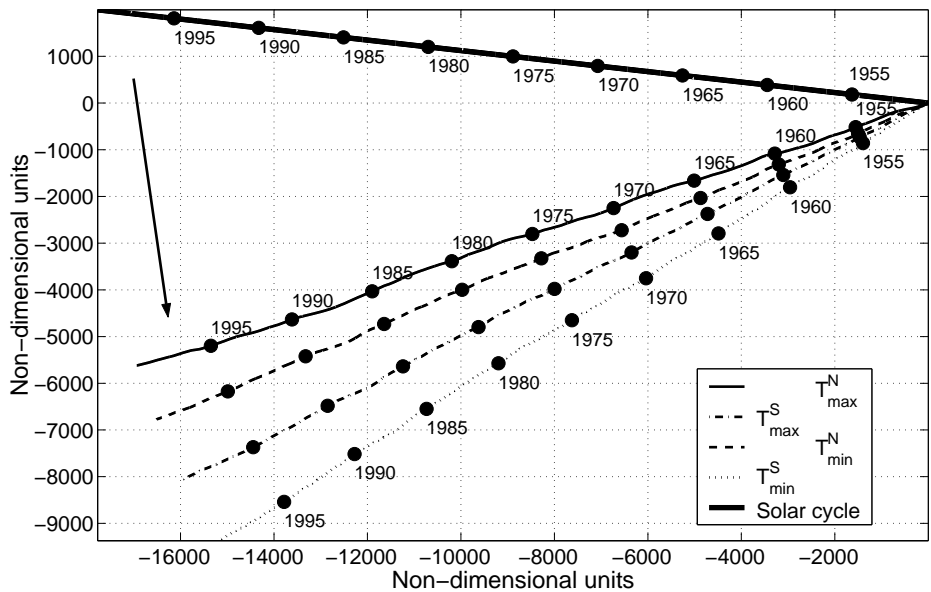}
 \caption{Phase cyclograms of the various temperature records and of the solar cycle. Abscissae: cumulative sum of the cosinusoidal coefficients. Ordinates: cumulative sum of the sinusoidal coefficients. The corresponding calendar years are indicated. The arrow points to the angle of increasing phase delay.}\label{cyclogram}
 \end{figure}

We present in figure \ref{cyclogram} the cyclograms of the four
signals $T_{max}^{N}$, $T_{max}^{S}$, $T_{min}^{N}$, and
$T_{min}^{S}$. In the same figure it is reported the phase
cyclogram that can be constructed from the rigorously periodic
solar cycle signal, which can be expressed as follows:
\begin{equation}
SC\left(t\right)=\cos\left(2\pi/\tau_0 t +\phi\right)
\end{equation}
where $\phi$ is such that for $t=171$ (corresponding to June
$21^{st}$) the argument of the cosine function is $0$. We observe
that in all four cases of the temperature data sets the cyclograms
are almost indistinguishable from straight lines, since the
$t-$dependent phase functions are essentially stationary. The
above mentioned average phase differences are the angles -
measured counter clock-wise - between the best straight line
estimates of the cyclograms considered.

We can interpret these results in physical terms as follows. On
one side, the lag and different amplitudes of the cycles of
maximum and minimum temperatures can be related to the different
impacts of changes of the two well-distinct processes of day solar
shortwave heating and night longwave cooling on the local
thermodynamic systems where measurements are taken, in terms of
relations to the thermal inertia. On the other side, larger scale
thermal inertia effects related to the different thermal
properties of sea and land provide a qualitative argument for the
differences in amplitude and phase of the station N and S cycles,
the main reason being that Northern Italy is more continental than
Southern Italy.

\section{Estimation of the significativity of the trends} \label{trend}
We have followed a Montecarlo approach in order to assess the
significativity of the computed trends for both the seasonal cycle
amplitude $A\left(t,2\pi/\tau_0\right)$ and phase
$\phi\left(t,2\pi/\tau_0 \right)$ of the four temperature records
analyzed.
\par Our procedure consists in adopting a null-hypothesis, so that
we assume that the considered quantity is a stationary
autoregressive signals of order $n$, which can in general be
expressed as:
\begin{equation}\label{autoreg}
w\left(t\right)=m+\sum_{k=1}^{n}c_k
w\left(t-k\right)+\eta\left(t\right)
\end{equation}
where $\eta$ is a white spectrum noise with variance
$\sigma_{\eta}$. We estimate for the considered quantity the
optimal order $n$, as well as the optimal values of the relevant
parameters $m$, $\{c_k\}$, and $\sigma_{\eta}$ of the
corresponding autoregressive process (\ref{autoreg}). This can be
performed, \textit{e.g.} using a suitable MATLAB$^\copyright$
routine \citep{Neumaier2001,Schneider2001}. We wish to emphasize
that the routine \citep{Neumaier2001,Schneider2001} allows to
estimate the optimal $n$ with either the Schwarz's Bayesian
criterion  \citep{Schwarz1978} (henceforth, SBC) or the logarithm
of Akaike's final prediction error \citep{Akaike1971} (henceforth,
FPE). The former approach gives consistently in all cases analyzed
smaller values for $n$. It has been shown in a simulation study
that SBC is the most efficient in selecting the correct model
order compared to other selection methods, among which FPE
\citep{Lutkepohl1985}. Since we are interested in robust
estimates, we have generally adopted the SBC.

We then perform a Montecarlo experiment by running several times
the autoregressive system having the previously obtained optimal
parameters and compute the statistics of the outputs. In such
simulations, the initial conditions are essentially not
 relevant in statistical terms. Anyway, in order to eliminate transient
effects and consider statistical equilibrium conditions, granted
by the stationarity of the process, we do not consider the first
1000 time steps.

This approach allows us to obtain an estimate of the standard
deviation of the trend. This analysis gives in all cases a
negative result, \textit{i.e.} we obtain non
statistically-significant trends. The $95\%$ confidence intervals
consistent with the null trend hypothesis are shown in tables
\ref{amptable} and \ref{phasetable} for the amplitude and phase
functions, respectively.

\section{Notes on the de-seasonalized data}\label{deseason}

\begin{figure}[t]
\centering
 \subfigure[$T_{max}^{N}(t)-\overline{\Sigma}\left(t,2\pi/\tau_0\right) \left\{T_{max}^{N}\right\}$]
   {\includegraphics[angle=270,width=0.45\textwidth]{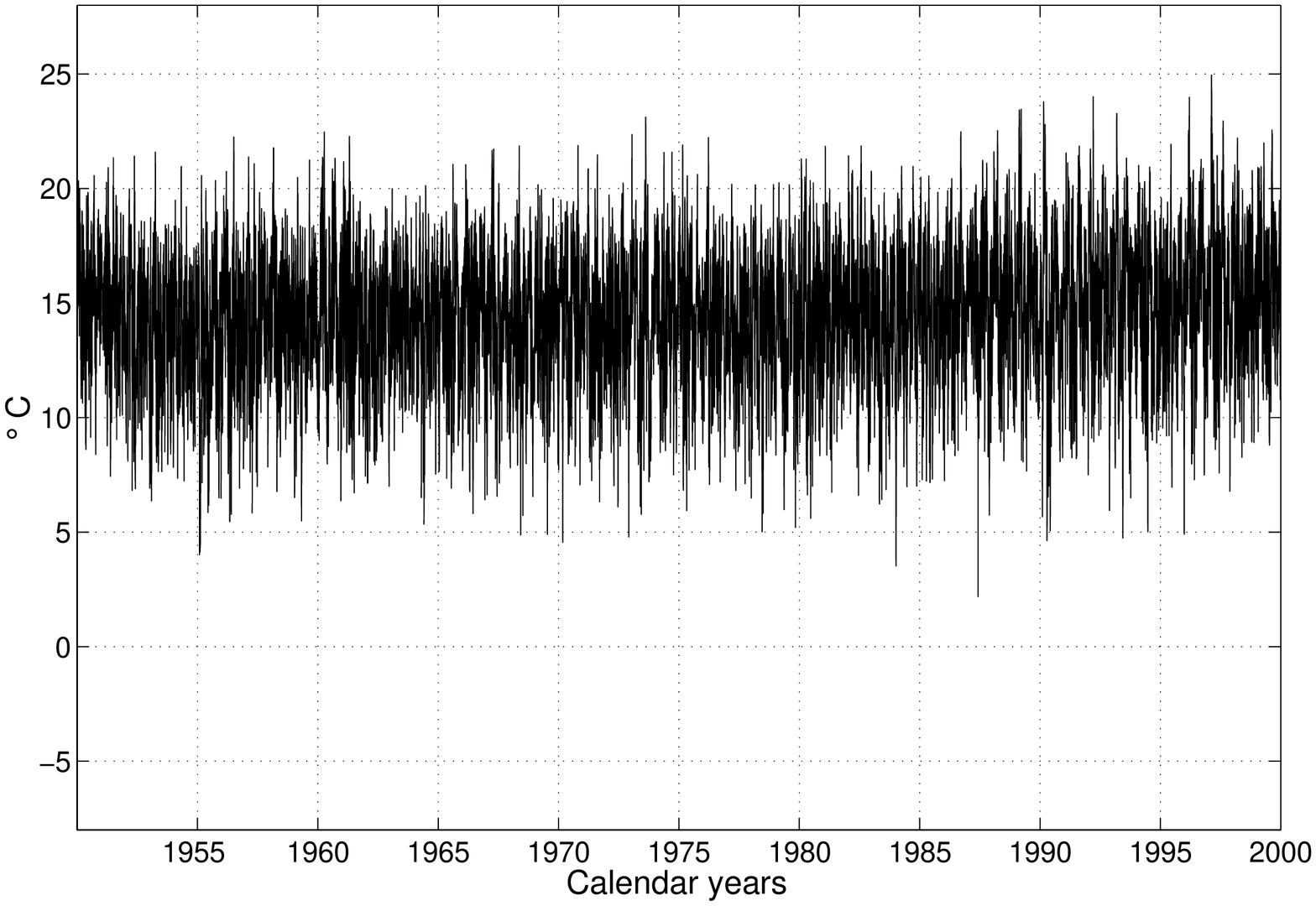}}
 \hspace{5mm}
 \subfigure[$T_{max}^{S}(t)-\overline{\Sigma}\left(t,2\pi/\tau_0\right)\left\{T_{max}^{S}\right\}$]
   {\includegraphics[angle=270,width=0.45\textwidth]{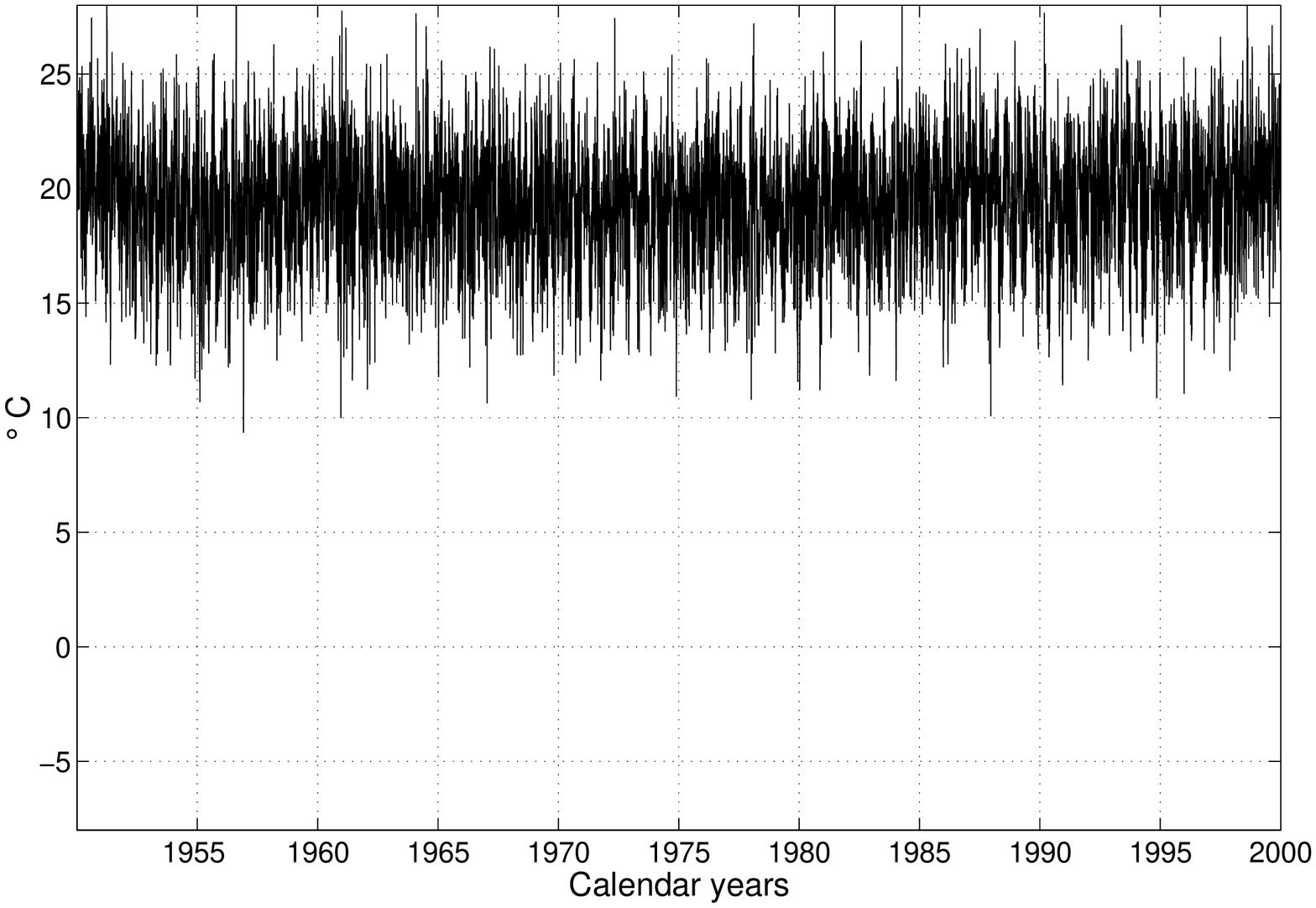}}\\
 \subfigure[$T_{min}^{N}(t)-\overline{\Sigma}\left(t,2\pi/\tau_0\right)\left\{ T_{min}^{N} \right\}$]
   {\includegraphics[angle=270,width=0.45\textwidth]{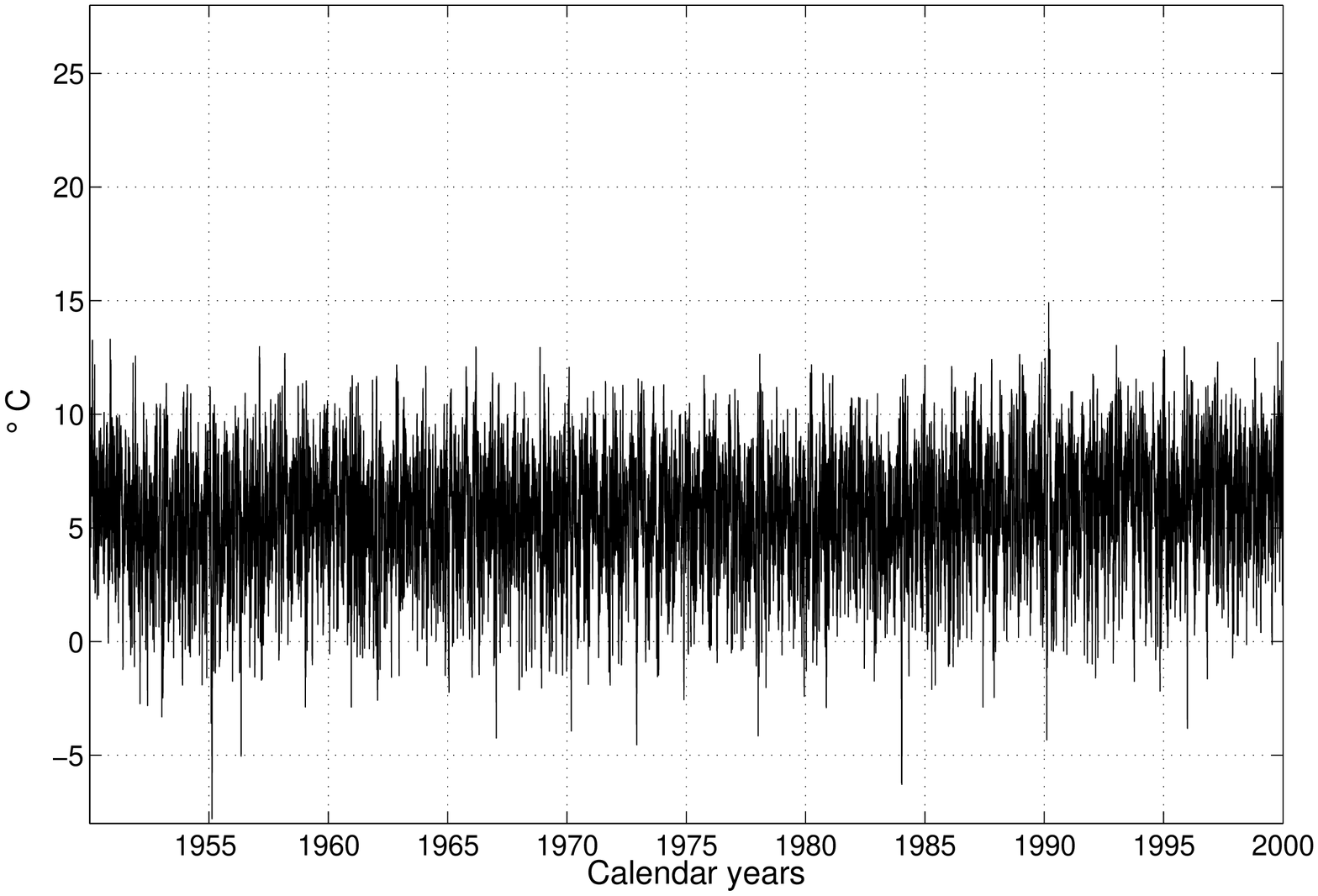}}
 \hspace{5mm}
 \subfigure[$T_{min}^{S}(t)-\overline{\Sigma}\left(t,2\pi/\tau_0\right) \left\{ T_{min}^{S}\right\}$]
   {\includegraphics[angle=270,width=0.45\textwidth]{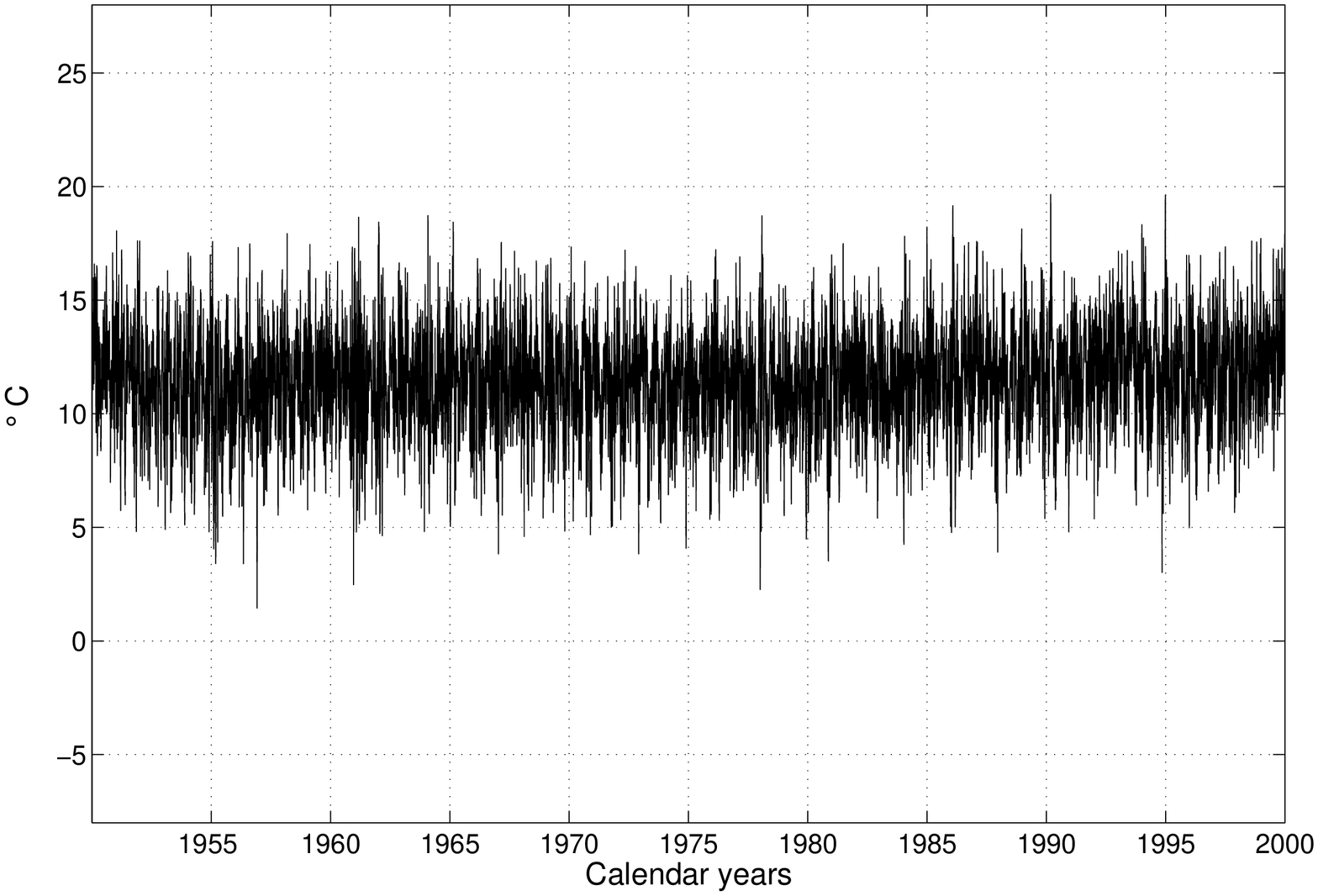}}
 \caption{De-seasonalized maximum and minimum temperature records of station N and station S.}\label{nocyclefigure}
 \end{figure}

In figures \ref{nocyclefigure} we present the data sets obtained
by subtracting the computed seasonal cycles to the corresponding
temperature records. These data have been fitted with
autoregressive models, whose order and parameters have also been
estimated with suitable software \citep{Neumaier2001,
Schneider2001}. The main statistical properties of these data are
presented in table \ref{datatable}. In all cases the estimated
optimal value of the autoregressive order, where the SBC has been
adopted, is between 3 and 5, which closely resembles the
characteristic time scale of the mid-latitude cyclones. The
variability of the subtracted signal, which can be estimated by
the value of the corresponding standard deviation, is larger for
the variables referring to station N, and largest for
$T_{max}^{N}$. This might be related to the fact that
mid-latitudes baroclinic weather disturbances are stronger in
Northern Italy, while the Southern Italy climate is less
influenced by such meteorological features. The climate of
Southern Italy might depend more on the strength and position of
the Hadley cell, which has a less pronounced short-time
variability.

Comparing the last two columns of table \ref{datatable}, we see
that in all cases the variance of the de-seasonalized signal is
smaller by about $5\%$ than the signal obtained by erasing the
$1y^{-1}$ frequency component computed over all the spectrum. This
implies that the local estimate of the seasonal cycle can explain
a larger fraction of the total variance of the signal than the
rigorously periodic seasonal signal obtained with DFT.

We underline that a correct evaluation of the seasonal signal is
of outstanding importance for a correct approach to the problem of
determining the extremes of a given climate record. In the case of
the data sets under investigation in this work, a thorough
analysis of the extremes will be shortly presented in a future
publication.

\section{Conclusions}\label{concl}
In this work we have analyzed the data sets covering the last 50
years of daily maximum and minimum temperature which are
representative of the Northern and of the Southern Italy
temperature fields, respectively.

We have analyzed the seasonal cycle with the technique of
cyclograms, which allows to find at each time the
quasi-instantaneous best estimate of the annual component of the
record. The resulting seasonal signal is not strictly periodic,
since at each time the  estimates of phase and amplitude change
slightly.

It is important to underline that such an approach is viable
because our signal obeys the narrow band approximation,
\textit{i.e.} the spectrum of the signal has a strong, narrow peak
for the annual cyclic component. If, on the contrary, the signal
is characterized by a broad spectral feature comprising the $1
y^{-1}$ frequency component, it is a mathematical nonsense to
investigate whether the seasonal cycle is changing. In such a case
the seasonal cycle is just \textit{not defined}, because several
contiguous spectral components having different frequencies and
shifting phase differences give contributions of comparable
importance.

In all cases analyzed, the time-dependent estimates of amplitude
and phase of the seasonal cycles do not show any statistically
significant trend in the time frame considered. Moreover, in each
case the average value of the estimates closely resemble the
amplitude and phase of the 1 year frequency sinusoidal signal
resulting from the Fourier analysis of the whole data set.
Succinctly, \textit{seasons seem to have not changed} in their
annual evolution.

In general, the amplitude of the maximum temperature seasonal
cycle is larger than that of the minimum temperature, and seasonal
cycles of station N are larger than those of station S. In terms
of phase, we observe that in general the minimum temperature
seasonal cycle lags behind the maximum temperature seasonal cycle,
and that the seasonal cycles of the station S lag behind the
corresponding cycles of the station N. All seasonal cycles lag
considerably behind the solar cycle.

On one side, thermal inertia effects related to the day/night
cycle explain the lag and different amplitudes of the cycles of
maximum and minimum temperatures. On the other side, larger scale
thermal inertia effects related to the different thermal
properties of sea and land provide a qualitative argument for the
differences in amplitude and phase of the station N and S cycles.
We underline that Northern Italy is more continental than Southern
Italy. The data support that two effects, which we have physically
referred to the the North-South and maximum-minimum (or day-night)
asymmetries, sum up linearly both for phase and amplitude of the
seasonal signals.

The data obtained by subtracting from the signal the corresponding
seasonal cycle have been fitted as autoregressive systems, whose
order and parameters have also been estimated with suitable
software. In all cases the optimal value of the autoregressive
order is between 3 and 5 (which is expressed in term of days)
which closely resembles the characteristic time scale of the
mid-latitude cyclones. The variability of the subtracted signal,
which can be estimated by the value of the corresponding standard
deviation, is larger for the northern variables, and largest for
$T_{max}^{N}$.

This might suggest that the climate of Northern Italy is strongly
driven, in statistical sense, by the southern portions of the
storm-track Atlantic eddies, while we might guess that the
northernmost branch of the Hadley cell plays a very relevant role
for the climate of Southern Italy. In future work it would be
possible to test such hypothesis by correlating the seasonal
signal of the temperature records here analyzed with the seasonal
signal of suitably defined indicators of storm-track activity and
meridional circulation.


Finally, we wish to emphasize two major limitations of the present
work with the perspective of providing hints for future research.
We wish to underline that if on one side the surface temperature
is a very relevant quantity in terms of influence on the
biosphere, including human activities, on the other side it is not
the most relevant quantity in terms of representing schematically
the thermodynamic properties of the system. As well known, a
measure of the average tropospheric temperature is much more
relevant in this sense \citep{PeixotoandOort1992}. Therefore, a
more physically sensitive approach would be considering the
records of the whole vertical temperature profile. Obviously, this
requires the availability of long and reliable radiosonde records.

Moreover, it is important to note that, when considering a limited
area, the direct solar forcing is \textit{not} the only relevant
forcing, since air advection at all levels from nearby areas plays
a fundamental role in determining the state of the system under
consideration. This is of special significance for areas, such the
Mediterranean basin or \textit{a fortiori} Italy, which do not
have a strong \textit{endogeneous} climate mode, as occurs in the
case of the Indian Monsoon area or Siberia, and are characterized
by an essentially residual climate.

Therefore, it would be important to consider in future analyses
the estimates of the convergence of thermal fluxes obtained from
the available reanalyses. It is important to note that especially
in the case of relatively small and elongated territories such as
Italy, the resolution of the data becomes of critical relevance.

\section*{Acknowledgments} We wish to thank for technical and
scientific help Mara Felici and Michele Brunetti. We also wish to
thank the AM for having provided most of the data used in this
work.  We also wish to thank Dr. Neumaier and Dr. Schneider for
making their MATLAB$^\copyright$ code freely available
 at the URL \\\texttt{http://www.gps.caltech.edu/$\sim$tapio/arfit/}

\clearpage
\newpage
\listoftables
\clearpage
\newpage
\listoffigures



\clearpage
\begin{table}
 \centering
\begin{minipage}[c]{\textwidth}
\begin{tabular}{|l|c|c|c|c|}
  \hline
  Variable & $\langle \textrm{Variable}\rangle$ & $2\sigma(Variable)$ & Estimated Trend & $2\sigma_{Trend}$  \\
  \hline
  $A\left\{T_{max}^{N}\right\}$ & $10.19$ $^{\circ}C$ & $1.34$ $^{\circ}C$  & [${0.002 ^{\circ}C/y}$] & $0.02$ $^{\circ}C/y$ \\
  $A\left\{T_{max}^{S}\right\}$ & $8.79$ $^{\circ}C$ & $1.27$ $^{\circ}C$  & [${0.006 ^{\circ}C/y}$] & $0.02$ $^{\circ}C/y$  \\
  $A\left\{T_{min}^{N}\right\}$ & $8.65$ $^{\circ}C$ & $1.30$ $^{\circ}C$  & [${0.004 ^{\circ}C/y}$] & $0.02$ $^{\circ}C/y$ \\
  $A\left\{T_{min}^{S}\right\}$ & $7.33$ $^{\circ}C$ & $1.28$ $^{\circ}C$  & [${0.010 ^{\circ}C/y}$] & $0.02$ $^{\circ}C/y$  \\
  \hline
\end{tabular}
  \caption{Statistical analysis of the amplitude of the seasonal cycle of the 4 variables considered. Estimated trends are not
statistically significant and the values are indicated between
brackets.}\label{amptable}
\end{minipage}
\end{table}
\newpage
\begin{table}
 \centering
\begin{minipage}[c]{\textwidth}
\begin{tabular}{|l|c|c|c|c|}
  \hline
  Variable & $\langle \textrm{Variable}\rangle$ & $2\sigma(Variable)$ & Estimated Trend & $2\sigma_{Trend}$  \\
  \hline
  $\phi\left\{T_{max}^{N}\right\}$ & $2.82$ Rad & $0.14$ Rad & [$0.0005$ Rad $y^{-1}$]   & $0.002$ Rad $y^{-1}$ \\
  $\phi\left\{T_{max}^{S}\right\}$ & $2.67$ Rad & $0.11$ Rad & [$-0.0002$ Rad $y^{-1}$] &  $0.002$ Rad $y^{-1}$ \\
  $\phi\left\{T_{min}^{N}\right\}$ & $2.75$ Rad & $0.12$ Rad & [$-0.00008$ Rad $y^{-1}$] & $0.002$ Rad $y^{-1}$ \\
  $\phi\left\{T_{min}^{S}\right\}$ & $2.60$ Rad & $0.11$ Rad & [$-0.0002$ Rad $y^{-1}$]  & $0.002$ Rad $y^{-1}$ \\
\hline
\end{tabular}
\caption{Statistical analysis of the phase of the seasonal cycle
of the 4 variables considered. Estimated trends are not
statistically significant and the values are indicated between
brackets.}\label{phasetable}
\end{minipage}
\end{table}
\newpage
\begin{table}
 \centering
\begin{minipage}[c]{\textwidth}
\begin{tabular}{|c|c|c|c|}
  \hline
  Variable & $\langle \textrm{Variable}\rangle$ & $2\sigma(Variable)$ &  $2\sigma(Variable)$[DFT] \\
  \hline
  $T_{max}^{N}-\overline{\Sigma}\left\{T_{max}^{N}\right\}$  & $14.5$ $^{\circ}C$ & $5.7$ $^{\circ}C$ & $5.9$ $^{\circ}C$ \\
  $T_{max}^{S}-\overline{\Sigma}\left\{T_{max}^{S}\right\}$ & $19.5$ $^{\circ}C$ & $4.8$ $^{\circ}C$ & $4.9$ $^{\circ}C$ \\
  $T_{min}^{N}-\overline{\Sigma}\left\{T_{min}^{N}\right\}$ & $5.7$ $^{\circ}C$ & $5.3$ $^{\circ}C$  & $5.4$ $^{\circ}C$  \\
  $T_{min}^{S}-\overline{\Sigma}\left\{T_{min}^{S}\right\}$ & $11.4$ $^{\circ}C$ & $4.2$ $^{\circ}C$ & $4.3$ $^{\circ}C$  \\
  \hline
\end{tabular}
  \caption{Statistical analysis of the de-seasonalized signal obtained with the cyclograms approach as compared to the results obtained with a conventional DFT approach.}  \label{datatable}
\end{minipage}
\end{table}

\clearpage
\newpage


\clearpage
\newpage
\begin{thebibliography}{15}
\expandafter\ifx\csname
natexlab\endcsname\relax\def\natexlab#1{#1}\fi
\expandafter\ifx\csname url\endcsname\relax
  \def\url#1{{\tt #1}}\fi
\expandafter\ifx\csname
urlprefix\endcsname\relax\def\urlprefix{URL }\fi
\expandafter\ifx\csname
doiprefix\endcsname\relax\def\doiprefix{doi:}\fi

\bibitem[{Akaike(1971)}]{Akaike1971}
Akaike, H., 1971: Autoregressive model fitting for control. {\it
Ann. Inst.
  Statist. Math.\/}, {\bf 23}, 163--180.

\bibitem[{Attolini et~al.(1984)Attolini, Cecchini, and Galli}]{Attolini1984}
Attolini, M.~R., S.~Cecchini, and M.~Galli, 1984: {\it Il Nuovo
Cimento C\/},
  {\bf 7}, 245.

\bibitem[{Attolini et~al.(1985)Attolini, Galli, and Castagnoli}]{Attolini1985}
Attolini, M.~R., M.~Galli, and G.~C. Castagnoli, 1985: On the
rz-sunspot
  relative number variations. {\it Solar Physics\/}, {\bf 96}, 391.

\bibitem[{Attolini et~al.(1989)Attolini, Galli, Nanni, and
  Povinec}]{Attolini1989}
Attolini, M.~R., M.~Galli, T.~Nanni, and P.~Povinec, 1989: A
cyclogram analysis
  of te bratislava 14c tree-ring record during the last century. {\it
  Radiocarbon\/}, {\bf 31}, 839--845.

\bibitem[{Brunetti et~al.(2001)Brunetti, Colacino, Maugeri, and
  Nanni}]{Brunetti2001}
Brunetti, M., M.~Colacino, M.~Maugeri, and T.~Nanni, 2001: Trends
in the daily
  intensity of precipitation in {Italy} from 1951 to 1996. {\it Int. J.
  Clim.\/}, {\bf 21}, 299--316.

\bibitem[{Brunetti et~al.(2002)Brunetti, Maugeri, Nanni, and
  Navarra}]{Brunetti2002}
Brunetti, M., M.~Maugeri, T.~Nanni, and A.~Navarra, 2002: Droughts
and extreme
  events in regional daily italian precipitation series. {\it Int. J. Clim.\/},
  {\bf 22}, 543--558.

\bibitem[{Galli(1988)}]{Galli1988}
Galli, M.: 1988, Time series analysis with power spectrum and
cyclograms. {\it
  Solar-Terrestrial Relationship and the earth environment in the Last
  Millennia\/}, G.~C. Castagnoli, ed., North Holland, Amsterdam, volume XCV of
  {\it Proceedings of the International School of Physics Enrico Fermi\/}, 246.

\bibitem[{Luetkepohl(1985)}]{Lutkepohl1985}
Luetkepohl, H., 1985: Comparison of criteria for estimating the
order of a
  vector autoregressive process. {\it J. Time Ser. Anal.\/}, {\bf 6}, 35--52.

\bibitem[{Maugeri et~al.(2001)Maugeri, Bagnati, Brunetti, and
  Nanni}]{Maugeri2001}
Maugeri, M., Z.~Bagnati, M.~Brunetti, and T.~Nanni, 2001: Trends
in italian
  total cloud amount. {\it Geophys. Res. Lett.\/}, {\bf 28}, 4551--4554.

\bibitem[{Maugeri et~al.(2003)Maugeri, Brunetti, Monti, and
  Nanni}]{Maugeri2003}
Maugeri, M., M.~Brunetti, F.~Monti, and T.~Nanni, 2003: The
italian air force
  sea level pressure data set (1951-2000). {\it Il Nuovo Cimento C\/}, {\bf
  26}, 453--467.

\bibitem[{Neumaier and Schneider(2001)}]{Neumaier2001}
Neumaier, A. and T.~Schneider, 2001: Estimation of parameters and
eigenmodes of
  multivariate autoregressive models. {\it ACM Trans. Math. Softw.\/}, {\bf
  27}, 27–57.

\bibitem[{Peixoto and Oort(1992)}]{PeixotoandOort1992}
Peixoto, A. and B.~Oort, 1992: {\it Physics of Climate\/}.
American Institute
  of Physics, Washington.

\bibitem[{Schneider and Neumaier(2001)}]{Schneider2001}
Schneider, T. and A.~Neumaier, 2001: Algorithm 808: Arfit - a
matlab package
  for the estimation of parameters and eigenmodes of multivariate
  autoregressive models. {\it ACM Trans. Math. Softw.\/}, {\bf 27}, 58–65.

\bibitem[{Schwarz(1978)}]{Schwarz1978}
Schwarz, G., 1978: Estimating the dimension of a model. {\it Ann.
Statist.\/},
  {\bf 6}, 461--464.

\bibitem[{Vinnikov et~al.(2003)Vinnikov, Robock, Grody, and
  Basist}]{Vinnikov2003}
Vinnikov, K.~Y., A.~Robock, N.~C. Grody, and A.~Basist, 2003:
Analysis of
  diurnal and seasonal cycles and trends in climatic records with arbitrary
  observation times. {\it Geophys. Res. Lett.\/}, {\bf 31}, L06205,
  {DOI}:10.1029/2003GL019196.

\end{thebibliography}
\end{document}